\documentclass[twocolumn,floats,floatfix,showpacs,prd,superscriptaddress,nofootinbib]{revtex4-1}

\usepackage{graphicx,epsfig}
\usepackage{amssymb,amsmath,amsthm,amsfonts}
\usepackage{bm} 

\usepackage[inline]{enumitem}
\usepackage[linktocpage]{hyperref}
\usepackage[caption=false]{subfig}
\usepackage[usenames]{color}
\usepackage{url}

\renewcommand{\vec}[1]{\boldsymbol{#1}}
\newcommand{\Ord}[2]{\mathcal O \left(#1^{#2}\right)}

\newcommand{\normfct}[1]{\left\lvert #1 \right\rvert}

\def\lean{\textsc{Lean}}
\def\canuda{\textsc{Canuda}}
\def\ETK{\textsc{Einstein Toolkit}}

\def\dif{\textrm{d}}
\def\p{\partial}

\def\rp{r_{+}}
\def\rmm{r_{-}}
\def\rex{R_{\rm{ex}}}

\def\guGR{g^{\rm{GR}}}
\def\gdZ{g_{(0)}}
\def\guZ{g^{(0)}}

\def\huI{h^{(1)}}

\def\PhiZ{\Phi^{(0)}}
\def\PhiI{\Phi^{(1)}}
\def\PhiII{\Phi^{(2)}}
\def\tM{t_{\rm{merger}}}                
\def\DPGW{\Delta\phi_{\rm{GW}}}         
\def\DPdet{\Delta\phi_{\rm{det}}}       

\def\alp3G{\alpha_{\rm{GB,3G}}}

\def\Lie{\mathcal{L}}

\def\F{\mathcal{F}}
\def\G{\mathcal{G}}
\def\H{\mathcal{H}}
\def\M{\mathcal{M}}

\def\P{\hat{\mu}}
\def\R{\mathcal{R}}

\begin{document}
{\hfill KCL-PH-TH/2018-59}

\title{Black holes and binary mergers in scalar Gauss--Bonnet gravity: scalar 
field dynamics}

\author{Helvi Witek}\email{helvi.witek@kcl.ac.uk}
\affiliation{Department of Physics, King's College London, Strand, London, WC2R 
2LS, United Kingdom}
\affiliation{Departament de F\'{i}sica Qu\`{a}ntica i Astrof\'{i}sica 
\& Institut de Ci\`{e}ncies del Cosmos (ICCUB),
Universitat de Barcelona, Mart\'{i} i Franqu\`{e}s 1, E-08028 Barcelona,
Spain}

\author{Leonardo Gualtieri}\email{leonardo.gualtieri@roma1.infn.it}
\affiliation{Dipartimento di Fisica, ``Sapienza'' Universit\'{a} di Roma \& Sezione INFN Roma 1,
P.A. Moro 5, 00185, Roma, Italy}

\author{Paolo Pani}\email{paolo.pani@roma1.infn.it}
\affiliation{Dipartimento di Fisica, ``Sapienza'' Universit\'{a} di Roma \& Sezione INFN Roma 1,
P.A. Moro 5, 00185, Roma, Italy}

\author{Thomas P.~Sotiriou}\email{Thomas.Sotiriou@nottingham.ac.uk}
\affiliation{School of Mathematical Sciences \& School of Physics and Astronomy, University of Nottingham,
University Park, Nottingham, NG7 2RD, UK
}

\begin{abstract}
We study the nonlinear dynamics of black holes that carry scalar hair and binaries composed of such black holes. The
scalar hair is due to a linear or exponential coupling between the scalar and the Gauss--Bonnet invariant. We work
perturbatively in the coupling constant of that interaction but nonperturbatively in the fields. We first consider the
dynamical formation of hair for isolated black holes of arbitrary spin and determine the final state. This also allows
us to compute for the first time the scalar quasinormal modes of rotating black holes in the presence of this
coupling. We then study the evolution of nonspinning black-hole binaries with various mass ratios and produce the first
scalar waveform for a coalescence. An estimate of the energy loss in scalar radiation and the effect this has on orbital
dynamics and the phase of the GWs (entering at quadratic order in the coupling) shows that 
GW detections can set the most stringent constraint to date on theories that exhibit a coupling between a scalar field
and the Gauss--Bonnet invariant.
\end{abstract}

\maketitle
\tableofcontents

\section{Introduction}\label{sec:intro}
The history of alternative theories of gravity is almost as old as that of general relativity~(GR) itself (see,
e.g.,~\cite{Will:2014kxa,Berti:2015itd}). For more than a century, each astrophysical revolution and the corresponding
observational opportunity led to a new milestone test of gravity. Einstein's theory has remained consistent with
observations (although dark matter and dark energy may be considered as indication to the contrary), while several
modified theories of gravity have been strongly constrained or ruled
out~\cite{Will:2014kxa,Yunes:2013dva,Berti:2015itd}.

The recent gravitational-wave~(GW) revolution provides yet another opportunity to test gravity in a new regime: the
highly-dynamical, strong-curvature regime probed by black holes~(BHs), compact objects, and binaries
thereof~\cite{Yagi:2016jml,Berti:2018cxi,Barack:2018yly}. While the recent GW events are all consistent with
GR~\cite{Yunes:2016jcc,TheLIGOScientific:2016src}, the constraints one can extract on alternative theories are rather
weak \cite{Yunes:2016jcc}, due to the lack of complete waveforms that correspond to binary evolution and mergers in
these theories.
Obtaining such waveforms is necessary to go beyond performing null tests of GR. Using theory-specific waveforms could
constrain the corresponding theory to unprecedented levels, or uncover new effects, using data that is already
available. Moreover, our ability to probe the highly nonlinear regime of gravity will improve further when LIGO/Virgo
detections will perform routinely at design sensitivity, and when future instruments such as third-generation
ground-based interferometers~\cite{Sathyaprakash:2012jk,Dwyer:2014fpa} and the future space mission
LISA~\cite{Audley:2017drz} will become operational.

Motivation for testing gravity are manifold~\cite{Will:2014kxa,Yunes:2013dva,Berti:2015itd,Barack:2018yly}, but arguably
the most pressing one is of fundamental nature: finding an underlying, consistent description of quantum gravity is
still the ``holy grail'' in modern physics.  GR itself fails at this task --~e.g., it is non renormalizable~-- and is
believed to be the leading-order manifestation of a more fundamental (possibly quantum) theory.
Remarkably, it has been shown~\cite{Stelle:1976gc} that including
terms  quadratic in the curvature in  the gravitational action can render the theory renormalizable 
(such terms also arise in the low-energy limit of string theories~\cite{Gross:1986mw}).

Uncovering a deviation from the predictions of GR would provide the first experimental insight into this fundamental
theory. There is no indisputable argument suggesting that new physics should make an appearance in BH binaries --- the
curvature involved is not high enough by fundamental physics standards. Nonetheless, it is significantly higher than the
curvature scale tested by any other observation and experiment, and it is particularly hard to argue against fully
exploring data from a new regime.

Perhaps the most obvious manifestation of new physics would be the existence of a new field, the simplest of which is a
(massless) scalar field $\Phi$.
No-hair theorems imply that scalar fields might be hard to detect with BHs, as the latter cannot generally support
nontrivial scalar configurations~\cite{1970CMaPh..19..276C,Hawking:1972qk,Bekenstein:1995un,Sotiriou:2011dz,Hui:2012qt,
Sotiriou:2013qea,Sotiriou:2015pka}.
Indeed, if one focuses on actions (including a single scalar field and also polynomial terms in the curvature tensor)
that yield second-order field equations, and imposing shift symmetry, {\em i.e.}~symmetry under $\Phi\to
\Phi\,+\,$constant, it turns out that there is only one interaction term that can induce scalar hair in stationary,
asymptotically flat configurations~\cite{Sotiriou:2013qea}: $\Phi\,\R_{\rm{GB}} $, where
\begin{align}
\label{eq:GBInvariant}
\R_{\rm{GB}} = & ^{(4)}R^{2} - 4\, ^{(4)}R_{ab}\, ^{(4)}R^{ab} +\, ^{(4)}R_{abcd}\, ^{(4)}R^{abcd} \,,
\end{align}
is the Gauss--Bonnet invariant, $^{(4)}R_{abcd}$ is the 4-dimensional Riemann tensor, $^{(4)}R_{ab}$ the corresponding
Ricci tensor, and $^{(4)}R$ the Ricci scalar. On the other hand, effective actions arising from string
theory~\cite{Mignemi:1992nt,Kanti:1995vq,Charmousis:2014mia} include the exponential coupling $e^\Phi \R_{\rm{GB}} $.
This term is well-known to lead to BH hair~\cite{Kanti:1995vq}.

Motivated by the above, the scalar Gauss--Bonnet (sGB) action
\begin{align}
\label{eq:ActionEdGB}
S = & \frac{1}{16\pi} \int \dif^{4} x \sqrt{-g} \left[ 
          \,^{(4)}R
        - \frac{1}{2} \left( \nabla\Phi \right)^{2} 
        + 2 \alpha_{\rm{GB}} f\left(\Phi\right) \R_{\rm{GB}}
\right]
\,,
\end{align} 
 has received considerable attention in the strong-field regime. Here we have employed geometric units, such that
 $G=1=c$, and $\alpha_{\rm{GB}}$ is the dimensionful GB coupling constant.\footnote{\label{foot}We summarize relations
   between different conventions for the coupling constant that are common in the literature in
   Appendix~\ref{app:CouplingConventions}.  For instance, the values of $\sqrt{\lvert\alpha_{\rm{GB}}\rvert}$ in the
   notation of~\cite{Kanti:1995vq} and of this article differ from those of~\cite{Yagi:2011xp,Yagi:2012gp,Yunes:2016jcc}
   by a factor of $4 \sqrt[4]{\pi}$.} Choosing $f(\Phi)= \frac{1}{8}e^{\Phi}$ corresponds to ``Einstein-dilaton
 Gauss--Bonnet'' (EdGB) gravity~\cite{Mignemi:1992nt,
   Kanti:1995vq,Pani:2011gy,Yagi:2011xp,Yagi:2012gp,Stein:2013wza,Yunes:2013dva,Ayzenberg:2014aka,Maselli:2015tta,Kleihaus:2015aje,
   Blazquez-Salcedo:2016yka}, while we refer to the choice
 $f(\Phi)=\Phi/8$~\cite{Sotiriou:2013qea,Sotiriou:2014pfa,Benkel:2016kcq, Benkel:2016rlz} as ``shift-symmetric sGB''
 gravity.  Recently, other interesting choices of $f$ have been
 studied~\cite{Silva:2017uqg,Doneva:2017bvd,Antoniou:2017hxj,Blazquez-Salcedo:2018jnn}.

As we will see in more detail later, the field equations of sGB gravity, though second order, contain highly nonlinear
quadratic terms of those second derivatives. Similar PDE structures, e.g. in hydrodynamics, are known to lead to shocks.
Hence, this raises reasonable concerns about the predictivity of the theory in the strong-field regime.  Indeed, it has
been shown that sGB combined with a generalized harmonic gauge is not well-posed~\cite{Papallo:2017qvl, Papallo:2017ddx}
although this is a gauge dependent result and, hence,
not conclusive.  A potential cure was suggested in~\cite{Cayuso:2017iqc,Allwright:2018rut} and will be adressed in
future work.

These problems only arise if action~\eqref{eq:ActionEdGB} is taken at face value. In this article we take a different
route and treat the coupling between $\Phi$ and $\R_{\rm{GB}}$ as the leading-order term of a low-energy expansion and
$\alpha_{\rm{GB}}$, or more correctly the dimensionless ratio $\epsilon\equiv\alpha_{\rm{GB}}/l^2$ (where $l$ is some
reference length) as the control parameter of this expansion. Within this framework, the theory is known to
$\Ord{\epsilon}{}$ only, solutions that are not smooth in the limit $\alpha_{\rm{GB}}\to 0$ should be considered
spurious, and hence one can solve the equations perturbatively in the coupling. This effective-field theory inspired
approach has been popular in the literature when obtaining stationary
solutions~\cite{Pani:2011gy,Yagi:2011xp,Maselli:2015tta} 
and recently it has also been used in the context of dynamical
evolution~\cite{Benkel:2016kcq,Benkel:2016rlz,Okounkova:2017yby}.

In what follows we will study the dynamics of isolated BHs and BH binaries for sGB gravity, working perturbatively and
up to linear order in the coupling, and for the choices $f=\Phi/8, e^{\Phi}/8$. As we will argue below, within the
perturbative treatment and at this order, these two choices are actually identical. 
Moreover, they are also equivalent (modulo rescaling of the coupling $\alpha_{\rm{GB}}$) 
to any other choice of the function $f(\Phi)$, as long as $f'(0)\neq0$.

Dimensional analysis suggests that there are terms other than $f(\Phi) \R_{\rm{GB}}$ that could appear at
$\Ord{\epsilon}{}$ in an effective action (such as those including derivatives of the scalar field)
which we are implicitly neglecting. However, the fact that $\Phi \R_{\rm{GB}}$ is the only shift-symmetric term that
leads to BH hair suggests strongly that including ${\cal O}(\epsilon)$ terms that respect this symmetry will lead to
secondary corrections only. Indeed, we will demonstrate below that, within our setup, such corrections are ${\cal
  O}(\epsilon^2)$ at least.  Hence, we do consider our results as rather generic, at least at the qualitative
level. When shift symmetry is abandoned though, the scalar can acquire a mass which can lead to effects that our
analysis does not capture.

Our results on isolated BHs will provide new insights on the quasi-normal ringing of spinning BHs.  Perturbations of
spherically symmetric and static BHs in EdGB gravity have been studied
in~\cite{Blazquez-Salcedo:2016enn,Blazquez-Salcedo:2017txk}, 
where it has been shown that
gravitational and scalar perturbations are coupled\footnote{More precisely, only the polar gravitational sector is
  coupled to the scalar perturbations, while the axial gravitational sector is
  decoupled~\cite{Pani:2009wy,Blazquez-Salcedo:2016enn}.} and the quasinormal-mode~(QNM) spectrum contains two branches
of modes which have been respectively called {\it gravitational-led} and {\it scalar-led}, according to which sector
they reduce to when the coupling constant vanishes.
Toy models obtained with point particles plunging into hairy BHs in EdGB gravity suggest that both gravitational-led and
scalar-led QNMs are present in the post-merger GW ringdown signal~\cite{Blazquez-Salcedo:2016enn,Blazquez-Salcedo:2017txk}.

In the case of rotation, which is so far unexplored at perturbative level, it is reasonable to expect the same
qualitative structure for the QNM spectrum. An explicit perturbative computation is extremely challenging due to the
lack of (an extended version of) the Teukolsky formalism for BHs in EdGB gravity. Current estimates are based on the
assumption that, in the eikonal limit, the light-ring modes are related to the QNMs even in modified gravity
theories~\cite{Blazquez-Salcedo:2016enn,Glampedakis:2017dvb}, but in fact the coupling of the gravitational
perturbations with the scalar field breaks this analogy~\cite{Blazquez-Salcedo:2016enn}. Our nonperturbative analysis
circumvents the aforementioned issues.

Our numerical simulations of BH binaries give the first complete waveforms for scalar radiation, including the ringdown,
within the class of theories we are studying. Moreover, we 
%
estimate
the impact of the scalar radiation and the corresponding loss of energy on the binary evolution and its imprint on the
phase of GWs.  We use
%
this estimate
to argue that future GW detections can place stronger bounds on the
theory than the known ones.

Most of the current constraints on the coupling constant of sGB gravity theories, $\alpha_{\rm{GB}}$, have been derived
for EdGB gravity and shift-symmetric sGB gravity. In this case, observations on the orbital decay of low-mass X-ray
binaries lead to
$\sqrt{\lvert\alpha_{\rm{GB}}\rvert}\lesssim 10\,$km~\cite{Yagi:2012gp,Seymour:2018bce}~\footnote{Recall that the
  definition of the coupling adopted in ~\cite{Yagi:2012gp,Seymour:2018bce} differs by a factor $\sim5$ with ours; see
footnote~\ref{foot} and App.~\ref{app:CouplingConventions}.
}.
Slightly more stringent constraints could be set from the measurements of quasiperiodic oscillations in the X-ray
emission from accreting BHs, although the latter might be affected by large
systematics~\cite{Maselli:2014fca,Maselli:2017kic}. However, in several sGB theories a stronger bound can be obtained by
a theoretical constraint.
In these theories, a stationary BH solution with mass $M$ only exists if
\begin{align}
\label{theor_bound_edgb}
\epsilon \equiv & \frac{\alpha_{GB}}{4M^2}<\epsilon_{max}
\,,
\end{align}
where we have fixed the reference length $l$ to $2M$, and $M$ is the Arnowitt-Deser-Misner (ADM) mass which coincides
with the BH mass in the case of an isolated BH, and with the binary total mass in the case of BH binaries.  The
threshold $\epsilon_{max}\sim{\cal O}(1)$ depends on the specific sGB
theory~\cite{Kanti:1995vq,Sotiriou:2014pfa,Doneva:2017bvd,Antoniou:2017hxj, Silva:2017uqg}.  For spherically symmetric
BHs, $4\epsilon_{max} \simeq0.619$ in EdGB gravity~\cite{Pani:2009wy}, $4\epsilon_{max}\sim0.3$ in shift-symmetric sGB
gravity~\cite{Sotiriou:2014pfa}. When this bound is reached, a curvature singularity emerges from within the
horizon~\cite{Sotiriou:2014pfa}, and the solution does not describe a BH anymore. In the case of rotating BHs, the bound
becomes stronger (at least, in the case of EdGB gravity, where it has been shown~\cite{Kleihaus:2015aje} that
$\epsilon_{max}$ decreases as the spin increases, and vanishes at extremality). Therefore, the mere existence of a BH of
mass $M$ implies that $\alpha_{\rm{GB}}<4\epsilon_{max}M^2$. The lightest BH observed, J1655-40, has a mass
$M\simeq5.4\,M_\odot$, leading to $\sqrt{\alpha_{\rm{GB}}}<6.6\,$km. 
A similar upper bound ($\sqrt{\alpha_{\rm{GB}}}<5.4\,$km) was derived from the existence of neutron 
stars with $M\approx 2M_\odot$~\cite{Pani:2011xm}, but it is less robust, 
as it depends on the equation of state.
We remark that the above constraints do not apply to the class of theories 
found in \cite{Doneva:2017bvd,Antoniou:2017hxj,Silva:2017uqg}, which predict the 
existence of both Kerr BHs and ``scalarized'' BHs.

It should be noted that here we are not requiring the scalar field to have any cosmological significance, so we will not
discuss cosmological constraints in any detail. It is also worth emphasising the following: the coincident detection of
GWs and gamma rays from the binary neutron star merger GW170817 have constrained the speed of GWs to extremely high
accuracy~\cite{TheLIGOScientific:2017qsa,GBM:2017lvd}. This has in turn been used to place stringent bounds on a class
of theories that included sGB gravity~\cite{Lombriser:2015sxa,Lombriser:2016yzn,Ezquiaga:2017ekz,Sakstein:2017xjx,Baker:2017hug,Creminelli:2017sry} 
but these bounds rely on the assumption that the scalar field is to account for dark energy \cite{Tattersall:2018map}. Since
we are making no such assumption here, these bounds are inapplicable. Note that in asymptotically flat spacetimes, the speed of
GWs approaches unity asymptotically (see e.g.~\cite{Ayzenberg:2013wua}). 

In the present paper we report new GW-based constraints on the GB coupling from fully nonlinear simulations covering the
inspiral, merger and ringdown.  In particular, we compare estimates of the expected GW dephasing against that of current
GW detections and the forecast for future third-generation detectors.
For a BH binary like GW151226 with mass ratio $\sim1/2$ and total mass $\sim20M_{\odot}$, we find
$\sqrt{\lvert\alpha_{\rm{GB}}\rvert}\lesssim2.7$km -- 
%
comparable to
the theoretical constraints and about one
order of magnitude stronger than those based on the inspiral only~\cite{Yagi:2012gp,Yunes:2016jcc}.

\section{Setup}\label{sec:Setup}
\subsection{Action and field equations}\label{ssec:ActionAndEoMs}
Varying the action~\eqref{eq:ActionEdGB} with respect to the scalar field $\Phi$ and metric $g^{ab}$ yields their field
equations
\begin{subequations}
\label{eq:EoMsEdGBgeneral}
\begin{align}
\label{eq:EoMsEdGBgeneralSca}
\Box\Phi = & -2\alpha_{\rm{GB}} f'(\Phi) \R_{\rm{GB}}
\,,\\
\label{eq:EoMsEdGBgeneralTen}
G_{ab} = & - \alpha_{\rm{GB}}  \G_{ab}          + \frac{1}{2} T_{ab}
\,,
\end{align}
\end{subequations}
where $f' \equiv \dif f/\dif\Phi$.  $G_{ab} \equiv \, ^{(4)}R_{ab} - 1/2 g_{ab}\, ^{(4)}R$ is the Einstein tensor, the
canonical part of energy-momentum tensor for the scalar field is
\begin{align}
\label{eq:TmnSF}
T_{ab} = &  \nabla_{a}\Phi \nabla_{b} \Phi - \frac{1}{2} g_{ab} 
\nabla^{c}\Phi \nabla_{c}\Phi
\,,
\end{align}
and the modification due to the GB term is~\cite{Kanti:1995vq,Pani:2009wy}
\begin{align}
\label{eq:TmnGB}
\G^{\rm{GB}}_{ab} = & \frac{\delta \R_{\rm{GB}}}{\delta g^{ab}}
        =   2 g_{c(a} g_{b)d} \epsilon^{edfg} 
\nabla_{h}\left[^{\ast}R^{ch}{}_{fg} \nabla_{e} f \right]
\,\nonumber\\ = &
            2 g_{c(a} g_{b)d} \epsilon^{edfg} 
\nabla_{h}\left[^{\ast}R^{ch}{}_{fg} f' \nabla_{e} \Phi \right]
\,,
\end{align}
where $^{\ast}R^{ab}{}_{cd} = \epsilon^{abef}\, ^{(4)}R_{efcd}$ is the dual Riemann tensor, and $\epsilon^{abcd}$ is the
totally anti-symmetric Levi-Civita pseudo-tensor.

\subsection{Perturbative treatment in the coupling}\label{ssec:EFTapproach}

\subsubsection{Preliminaries}
%
Since we want to use a perturbative treatment in the coupling, we will assume that $\epsilon\ll1$ and formally expand
any tensor ${\bf{X}}$ as
\begin{align}
\label{eq:expansionGeneral}
{\bf{X}} = & \sum^{\infty}_{k=0} \frac{1}{k!} \epsilon^{k} {\bf{X}}^{(k)}\,.
\end{align}
In particular, the spacetime metric and the scalar field are expanded as
\begin{subequations}
\label{eq:expansionSFMetric}
\begin{align}
\label{eq:expansionSF}
\Phi   = & \sum^{\infty}_{k=0} \frac{1}{k!} \epsilon^{k} \Phi^{(k)}
\,,\\
\label{eq:expansionMetric}
g_{ab} = & \guZ_{ab} + \sum^{\infty}_{k=1} \frac{1}{k!} \epsilon^{k} 
h^{(k)}_{ab}
\,.
\end{align}
\end{subequations}
It should be stressed that this is not a weak-field expansion.
We raise indices of all tensorial quantities with $g^{(0)ab}$, e.g. we define
\begin{align}
h^{(k)ab} = & g^{(0){ac}}\,g^{(0){bd}}\, h^{(k)}_{cd}
\,,
\end{align}
and likewise for all other tensors.
Since $\epsilon$ is dimensionless, any tensor ${\bf{X}}^{(k)}$ has the same dimensions as the background tensor
${\bf{X}}^{(0)}$ -- for instance, the scalar field perturbations $\Phi^{(k)}$ are dimensionless, as is the background
scalar field, and so on.

\subsubsection{Field equations}
Applying the perturbative treatment to the field equations~\eqref{eq:EoMsEdGBgeneral} yields, to $\Ord{\epsilon}{}$,
\begin{subequations}
\label{eq:ExpansionFull}
\begin{align}
\label{eq:ExpansionFullEps0}
\epsilon^{0}:\quad & 
  G^{(0)}_{ab} =\frac{1}{2} T^{(0)}_{ab} 
\,,\quad
  \Box^{(0)} \PhiZ= 0
\,,\\
\label{eq:ExpansionFullEps1}
\epsilon^{1}:\quad & 
  G^{(1)}_{ab} = \frac{1}{2} T^{(1)}_{ab}- 4M^2 \G^{(0)}_{ab}   
\,,\\ &
  \Box^{(0)} \PhiI =-\Box^{(1)} \PhiZ -  8M^2f'_{(0)} \R_{\rm{GB}}^{(0)} 
\,,\nonumber
\end{align}
\end{subequations}
where $G^{(k)}_{ab}$, $\R^{(k)}_{\rm{GB}}$, $\G^{(k)}_{ab}$, $\Box^{(k)}$ and $T^{(k)}_{ab}$ refer to the $k$-th order
correction to the corresponding quantity.
The crucial feature of the equations above is that higher-curvature corrections at any given order always enter only as
source terms computed from the metric and the scalar field at lower order. Hence, at any given order the system of
partial differential equations can be made well-posed by an appropriate gauge choice or
reformulation~\cite{Salgado:2008xh,Wald:1984rg}.

\subsubsection{Zero-th order}
The zero-th order in the perturbative expansion, equivalent to taking the limit $\epsilon \rightarrow 0$ of
Eqs.~\eqref{eq:EoMsEdGBgeneral}, leads to Einstein's equations minimally coupled to a massless scalar field,
Eq.~\eqref{eq:ExpansionFullEps0}.
It has been shown that this system can be cast into a well-posed initial value formulation~\cite{Salgado:2008xh}, which
is a necessary condition for numerical stability.

Stationary, asymptotically flat BHs cannot carry hair if they satisfy
Eqs.~\eqref{eq:ExpansionFullEps0}~\cite{Hawking:1972qk}.  That is, they would be solutions of vacuum Einstein's
equations and any nontrivial initial scalar configuration would be shedded away.  One also expects that the scalar field
would not be excited in binaries composed of such BHs, as there is no scalar charge to begin with and the equations are
linear in the scalar. This suggests that, (at least) at late times, the solution to the zero-th order equation should be
of the form
\begin{align}
\label{eq:SolZerothOrder}
\left(\guZ_{ab},\PhiZ \right) = & \left( \guGR_{ab},0 \right)
\,,
\end{align}
where $\guGR_{ab}$ is a solution of the vacuum Einstein equations. However, there is a subtlety in this argument. As is
evident from Eqs.~\eqref{eq:ExpansionFullEps1}, $\PhiZ$ effectively sources the first-order (and subsequent order)
equations. Hence, a nontrivial initial $\PhiZ$ configuration could in principle leave some imprint on the
evolution. Though our expectation is that this effect would be rather small, we have not explored this in any detail.
Instead we focus on the late-time behaviour and we enforce $\PhiZ=0$. This choice will affect the form of the
first-order equations.

\subsubsection{First order}
Using the solution~\eqref{eq:SolZerothOrder} the first-order field equations~\eqref{eq:ExpansionFullEps1} reduce to
\begin{align}
\label{eq:ExpansionEps1Phi0}
G^{(1)}_{ab}     =  0
\,,& \quad
\Box^{(0)} \PhiI =  - 8M^2 f'_{(0)} \R_{\rm{GB}}^{(0)} 
\,,
\end{align}
where $f'_{(0)}=1/8$ for both EdGB and shift-symmetric sGB gravity, $\Box^{(0)}$ and $\R_{\rm{GB}}^{(0)}$ are,
respectively, the d'Alembertian and Gauss--Bonnet invariant evaluated from the background metric $\guZ_{ab}$.
$G^{(1)}_{ab}$ is the Einstein tensor acting on $\huI_{ab}$ with derivatives constructed from $\guZ_{ab}$.  Hence, the
metric itself is not deformed and it is safe to set $h^{(1)}_{ab}=0$.
As indicated in Eq.~\eqref{eq:ExpansionEps1Phi0} the scalar field $\PhiI$ is sourced by the curvature of the background
spacetime and, therefore, develops a nontrivial profile.
Then, the solution at ${\cal O}(\epsilon)$ is
\begin{align}
\label{eq:SolFirstOrder}
\left(\huI_{ab},\PhiI \right) = & \left( 0, \PhiI \right)
\,,
\end{align}
where $\PhiI$ can be solved for analytically in certain approximations discussed below, or numerically in the general
case.
Since Eqs.~\eqref{eq:ExpansionEps1Phi0} are the Einstein-scalar field equations sourced by tensors computed from
$(\guGR_{ab},0)$, they can be cast into a well-posed initial value formulation.

We remark that under the assumption $\Phi^{(0)}=0$ and for vanishing ordinary matter the scalar field is ${\cal
  O}(\epsilon)$ and the Ricci tensor is ${\cal O}(\epsilon^2)$, hence the GB invariant is equivalent to the Kretschmann
scalar up to ${\cal O}(\epsilon^4)$ terms.  Any other term which we have neglected in the action would be of the same
order or higher in the perturbation expansion (with the exception of the parity-violating Pontryagin density
$\epsilon_{abcd}R^{ab}_{~~ef}R^{fecd}$
leading to Chern-Simons gravity~\cite{Jackiw:2003pm,Alexander:2009tp,Delsate:2014hba}). Therefore, within our
perturbative approach and excluding parity violation, sGB gravity provides the most general theory with higher-order
curvature corrections.

\subsubsection{Second order}
Although we shall solve the field equations up to $\Ord{\epsilon}{}$ only, in order to assess the validity of our
perturbative approach and to estimate backreaction effects on the system's dynamics we inspect the field equations at
order $\Ord{\epsilon}{2}$.
Since $\Phi={\cal O}(\epsilon)$, its corrections to the metric appear at second 
order, as we now show.

{\noindent{\bf{Energy-momentum tensor:}}} When $\PhiZ=0$, one gets
\begin{align}
\label{eq:ExpansionTmnSF}
T^{(0)}_{ab} = & 0
\,,\quad
T^{(1)}_{ab} =   0
\,,
\end{align}
whereas the first nonvanishing contribution to the scalar stress-energy tensor reads
\begin{align}
\label{eq:ExpansionTmnSFEps2}
T^{(2)}_{ab} = & 2  \p_{a} \PhiI \p_{b} \PhiI - \guZ_{ab} 
\gdZ^{cd} \p_{c} \PhiI \p_{d} \PhiI \,.
\end{align}

{\noindent{\bf{Gauss--Bonnet correction and invariant:}}} 
Both quantities enter the field equations at a given order $\epsilon^{(k)}$ only as source terms, {\em i.e.}, computed
from lower-order terms.  Up to the linear level considered in~\eqref{eq:ExpansionFull} and inserting
solution~\eqref{eq:SolZerothOrder} we have 
(neglecting for simplicity the superscript $\,^{(4)}$
in front of the curvature tensor and its contractions)
\begin{subequations}
\begin{align}
\label{eq:ExpansionRGB}
\R^{(0)}_{\rm{GB}} = & R_{(0)}^{abcd} R^{(0)}_{abcd} - 4 R^{(0)}_{ab} 
R_{(0)}^{ab} + R_{(0)}^{2}
\,,\\
\label{eq:ExpansionTmnGB}
\G^{(0)}_{ab} = & 0
\,,
\end{align}
\end{subequations}
{\em i.e.}, the Gauss--Bonnet invariant depends only on the curvature of the background GR spacetime.

Instead, the GB correction does contribute to the energy-momentum content of the system at second order (which we will use
to estimate the GW dephasing), given by
\begin{align}
\label{eq:ExpansionTmnGBEps1}
\G^{(1)}_{ab} = & 2 \epsilon^{edfg} \guZ_{c(a} \guZ_{b)d} \nabla^{(0)}_{h} 
\left[f'_{(0)} \,^{\ast} R_{(0)}^{ch}{}_{fg} \p_{e} \PhiI 
        \right]
\,.
\end{align}
{\noindent{\bf{Field equations:}}} The field equations at order $\Ord{\epsilon}{2}$ read
\begin{subequations}
\label{eq:ExpansionEps2Phi0}
\begin{align}
\label{eq:ExpansionEps2Phi0Metric}
G^{(2)}_{ab}      = & - 8 M^2\G^{(1)}_{ab} + \frac{1}{2} T^{(2)}_{ab} 
\,,\\
\label{eq:ExpansionEps2Phi0SF}
\Box^{(0)} \PhiII = & - 8 M^2 f'_{(1)} \R^{(0)}_{\rm{GB}}
\,.
\end{align}
\end{subequations}
We remark that the second-order equations are different for EdGB gravity ($f'_{(1)}=\Phi^{(1)}/8$) and for
shift-symmetric sGB gravity ($f'_{(1)}=0$).

The right-hand side of~\eqref{eq:ExpansionEps2Phi0Metric} defines an effective energy-momentum tensor
\begin{align}
\label{eq:ExpansionEps2TmnEff}
T^{\rm{eff}}_{ab} = & T^{(2)}_{ab} - 16 M^2\G^{(1)}_{ab}
\,,
\end{align}
where $T^{(2)}_{ab}$ and $\G^{(1)}_{ab}$ are given in Eqs.~\eqref{eq:ExpansionTmnSFEps2}
and~\eqref{eq:ExpansionTmnGBEps1}, respectively.

We remark that, if one wishes to compute the first nonvanishing corrections to the metric components (including GW
emission), it is sufficient to solve the modified Einstein equations~\eqref{eq:ExpansionEps2Phi0Metric} with the
linear-order scalar field as an input, whereas the quadratic correction to the scalar field $\PhiII$ does not affect
the metric to leading order.

\subsubsection{Summary}
In the following, we set $f'_{(0)} = \tfrac{1}{8}$. This choice corresponds to the coupling functions
\begin{align}
\label{eq:FPhiExplicit}
f(\Phi)= & \frac{1}{8} e^{\Phi}
\,,\quad\textrm{or}\quad
f(\Phi)=\tfrac{1}{8}\Phi
\,,
\end{align} 
describing, respectively, EdGB gravity or shift-symmetric sGB gravity. However, we should stress that, within our
perturbative scheme, our results are far more general.  Indeed, based on the previous discussion, the choice of
$f'_{(0)}$ completely determines the form of the field equation for the scalar field up to order ${\cal O}(\epsilon)$,
cf.~Eqns.~\eqref{eq:ExpansionEps1Phi0}, and also the form of Einstein's equations at ${\cal O}(\epsilon^2)$,
cf.~Eq.~\eqref{eq:ExpansionEps2Phi0Metric}. Moreover, the precise value of $f'_{(0)}$ can be absorbed in
$\alpha_{\rm{GB}}$. Hence, one does not need to specify $f(\Phi)$ any further to fully determine the evolution to ${\cal
  O}(\epsilon)$ and to estimate how the scalar emission affects the gravitational waveform at ${\cal O}(\epsilon^2)$, as
we will do here.
  
Note however that theories for which $f'(0)=0$ are not covered by our analysis. It has been found recently that such
theories can yield interesting phenomena such as BH scalarization~\cite{Silva:2017uqg,Doneva:2017bvd,Antoniou:2017hxj},
so binary evolution in these theories deserves further consideration.

Summarizing,  the set of field equations (up to ${\cal O}(\epsilon)$) that we evolve numerically is
\begin{align}
\label{eq:EoMsEFTapproachEps1}
G^{(0)}_{ab} = & 0
\,,\quad
\Box^{(0)} \PhiI = - M^2\R^{(0)}_{\rm{GB}}
\,.
\end{align}
As discussed in Sec.~\ref{sec:EvolFormulation}, we evolve the scalar field simultaneously with the background, {\em i.e.}  GR
spacetime, that is set up either as a single rotating BH or a BH binary.

Unless needed for clarity, we will drop the superscripts $^{(0)},\,^{(1)}$ in the following, {\em i.e.} $\Phi\equiv\PhiI$ and
$g_{ab}\equiv\guZ_{ab}=\guGR_{ab}$.
%

\section{Formulation as time evolution problem}~\label{sec:EvolFormulation}
\subsection{Spacetime split}\label{ssec:SpacetimeSplit}
To write the field equations~\eqref{eq:EoMsEFTapproachEps1}
as a time evolution problem we perform a spacetime decomposition.
Specifically, we foliate the background spacetime $\left(\M,g_{ab}\right)$ into a set of spatial hypersurfaces
$\left(\Sigma_{t},\gamma_{ij}\right)$ labelled by a time parameter $t$ and with $3$-metric $\gamma_{ab}=g_{ab} +
n_{a}n_{b}$, where $n^{a}$ denotes the timelike unit vector normal to the hypersurface and is normalized such that
$n^{a}n_{a} = -1$.
The spatial metric defines a projection operator
\begin{align}
\label{eq:ProjOp}
\gamma^{a}{}_{b} = & \delta^{a}{}_{b} + n^{a} n_{b}
\,,
\end{align}
with $\gamma^{a}{}_{b} n^{b} = 0$ by construction.
The line element takes the form
\begin{align}
\label{eq:LineElement3p1}
\dif s^{2} = & g_{ab} \dif x^{a} \dif x^{b}
 \\
        = & - \left(\alpha^{2} - \beta^{k} \beta_{k} \right) \dif t^{2}
            + 2 \gamma_{ij} \beta^{i} \dif t \dif x^{j}
            + \gamma_{ij} \dif x^{i} \dif x^{j}
\,. \nonumber
\end{align}
where $\alpha$ and $\beta^{i}$ are, respectively, the lapse function and shift vector.
We denote the covariant derivative and Ricci tensor associated to the 3-metric $\gamma_{ij}$ as $D_{i}$ and $R_{ij}$,
respectively.
The extrinsic curvature is defined as
\begin{align}
\label{eq:DefKij}
K_{ab} = & - \gamma^{c}{}_{a} \gamma^{d}{}_{b} \nabla_{c} n_{d}
        = - \frac{1}{2} \Lie_{n} \gamma_{ab}
\,,
\end{align}
where 
$\Lie_{n}$ is the Lie derivative along $n^{a}$.
%

\subsection{Background spacetimes}~\label{ssec:BackgroundSpacetimes}
We consider two types of background spacetimes: 
\begin{enumerate*}[label={(\roman*)}]
\item Rotating BHs that will allow us to study the {\textit{dynamical}} formation of hairy BHs and their quasi-normal
  ringdown in the time domain.  This also allows us to benchmark our code at late times against the analytic solutions
  summarized in Appendix~\ref{ssec:BHsolsAna}. In this case we consider an isolated BH with total mass $M$ and
  dimensionless spin $\chi\equiv a/M = J/M^{2}$.
\item BH binaries that will enable us to explore, for the first time, the scalar excitation in sGB gravity induced by
  the strong-field dynamics of a coalescing compact binary in the background, and estimate its potential effect on the
  GW emission in this theory. In this case we consider a nonspinning binary with total mass $M$, mass ratio $q$ and
  initial separation $d/M$.
\end{enumerate*}
We briefly describe each of the settings below.

\subsubsection{Isolated rotating black holes}~\label{sssec:KerrInQIcoords}
We are interested in tracking the formation of scalar hair around a rotating BH.  As previously discussed, to linear
order in $\epsilon$ the background spacetime is a solution of vacuum Einstein's equations and, hence, the unique
stationary solution is the Kerr metric with mass $M$ and spin $J=a M$.
In Boyer-Lindquist coordinates $(t,r,\theta,\phi)$ its line element reads
\begin{align}
\dif s^{2} = & - \left( 1 - \frac{2 M r}{\Sigma} \right) \dif t^{2}
               - \frac{4 a M r \sin^{2}\theta }{\Sigma} \dif t \dif\phi
\nonumber \\ &
               + \frac{\Sigma}{\Delta} \dif r^{2}
               + \Sigma \dif\theta^{2}
               + \frac{\F}{\Sigma} \sin^{2}\theta \dif \phi^{2}
\,,
\end{align}
where the metric functions are defined as
\begin{subequations}
\begin{align}
\Delta = & r^{2} + a^{2} - 2 M r
       =   ( r - \rp ) ( r - \rmm )
\,,\\
\Sigma = & r^{2} + a^{2} \cos^{2}\theta
\,,\\
\F     = & \left( r^{2} + a^{2} \right)^{2} - \Delta a^{2} \sin^{2}\theta
\,,
\end{align}
\end{subequations}
and the inner and outer horizons are located at
\begin{align}
r_{\pm} =  M \pm \sqrt{M^{2} - a^{2} }
\,.
\end{align}

In light of the numerical evolutions of the scalar field dynamics it is convenient to introduce a quasi-isotropic radial
coordinate $R$~\cite{Liu:2009al,Okawa:2014nda}, such that
\begin{align}
\label{eq:QIradiusDef}
r =  R \left( 1 + \frac{\rp}{4 R} \right)^{2}
\,.
\end{align}
In the remainder of this section we use $r$ and $r_{\pm}$ purely as shorthand notation.
In these new coordinates the outer horizon is located at 
\begin{align}
R_{+} =  \frac{\rp}{4}
\,,
\end{align}
whereas the inner horizon is not part of the domain $R\in(0,\infty)$.
In contrast to the definition of Refs.~\cite{Brandt:1994ee,Brandt:1996si}, the transformation~\eqref{eq:QIradiusDef}
yields a finite (nonzero) horizon radius in the extremal limit $\lim_{a\rightarrow M} R_{+} = M/4$, thus allowing us to
set up highly spinning BH backgrounds.

After applying the coordinate transformation~\eqref{eq:QIradiusDef} and performing the spacetime
split~\eqref{eq:LineElement3p1}, the $3$-metric and gauge functions are given by
\begin{subequations}
\label{eq:MetricKQI}
\begin{align}
\label{eq:3MetricKQI}
\gamma^{K}_{ab} = & {\textrm{Diag}}\left[ \frac{\left( 4 R + \rp \right)^{2} 
\Sigma }{16 R^{3} \left(r-\rmm\right) },
        \Sigma, \frac{\F}{\Sigma} \sin^{2}\theta
        \right]
\,,\\
\label{eq:GaugeKQI}
\alpha_{K}      = & \pm \sqrt{\frac{\Delta \Sigma}{\F}}
\,,\quad
\beta^{a}_{K}   =   \left( 0, 0, - \frac{2 a M r}{\F} \right)
\,.
\end{align}
\end{subequations}
Here we have analytically continued the lapse function, and its positive (negative) sign corresponds to the exterior
(interior) region.
The nonvanishing components of the extrinsic curvature~\eqref{eq:DefKij}
read
\begin{align}
\label{eq:KijKQI}
K^{K}_{R\phi}      = & \frac{ \alpha_{K} a M r' 
\sin^{2}\theta}{\Delta\Sigma^{2}}
        \left[ 2 r^{2} \left(r^{2} + a^{2}\right) + \Sigma \left(r^{2} - 
a^{2}\right)
        \right]
\,,\nonumber\\
K^{K}_{\theta\phi} = & - 2 \alpha_{K} \frac{a^{3} M r \cos\theta 
\sin^{3}\theta}{\Sigma^{2}}
\,,
\end{align}
where $r' = \p_{R} r = 1 - \tfrac{\rp^{2}}{16R^{2}}$.

The Gauss--Bonnet invariant that sources the scalar field dynamics reduces to 
the Kretschmann scalar. For a single Kerr BH it is
\begin{align}
\label{eq:GBinvExplKQI}
\R_{\rm{GB}} = & \frac{48 M^{2}}{\Sigma^6} \left( 32 r^6 - 48 r^4 \Sigma + 18 
r^{2} \Sigma^{2} - \Sigma^{3} \right)
\,.
\end{align}
In practice, our (spatial) numerical domain is described in terms of Cartesian coordinates $X^{i}=\left\{x,y,z\right\}$.
Their relation to spherical coordinates $X^{a}=\left\{R,\theta,\varphi\right\}$ is given by
\begin{align}
\label{eq:TrafoSphVsCart}
x = & R \sin\theta \cos\varphi
\,,\,\,\,
y =  R \sin\theta \sin\varphi
\,,\,\,\,
z =  R \cos\theta
\,.
\end{align}
Applying the coordinate transformation~\eqref{eq:TrafoSphVsCart},
the spatial line element can be written explicitly as
\begin{align}
\dif l^{2} = & \gamma_{ij} \dif X^{i} \dif X^{j}
\nonumber\\ = &
        \psi^{4}_{0} \left[ \eta_{ij}  \dif X^{i} \dif X^{j} 
                + G \left( x\dif x + y\dif y + z\dif z \right)^{2}
\right. \nonumber \\ & \left.
                + a^{2} H \left( x\dif y - y\dif x \right)^{2}
        \right]
\,,
\end{align}
where $\eta_{ij}$ is the flat space metric and we introduced
\begin{align}
\psi^{4}_{0} = & \frac{\Sigma}{R^{2} }
\,,\quad
G = \frac{\rmm}{R^{2}(r-\rmm)}
\,,\quad
H = \frac{2 M r + \Sigma }{R^{2} \Sigma^{2}} 
\,.
\end{align}
The extrinsic curvature and shift vector transform according to
\begin{align}
K_{ij}    = & \Lambda^{a}{}_{i} \Lambda^{b}{}_{j} K^{K}_{ab}
\,,\quad
\beta^{i} =   \Lambda^{i}{}_{a} \beta^{a}_{K}
\,,
\end{align}
where $\Lambda^{i}{}_{a} = \p X^{i} / \p X^{a} $ is the Jordan matrix,
and $K^{K}_{ab}$ and $\beta^{a}_{K}$ are given in 
Eqs.~\eqref{eq:KijKQI}~and~\eqref{eq:GaugeKQI}, respectively.

\subsubsection{Black hole binaries}
Even within GR, the (near merger) two-body dynamics of BHs with comparable masses has to be solved numerically.  The
techniques are by now standard and regularly employed in GW source modelling, so we only give a brief summary of the
specific ingredients that we use and refer the interested reader to textbooks,
e.g.~\cite{Alcubierre:2008,Shibata:2015:NR:2904075,Baumgarte:2010:NRS:2019374}.
Specifically, we follow the ADM-York approach~\cite{Arnowitt:1962hi,York1979}.  Applying the spacetime split discussed
in Sec.~\ref{ssec:SpacetimeSplit} we can rewrite Einstein's equations -- a system of coupled partial differential
equations of mixed character -- as a set of elliptic-type constraint equations and a set of hyperbolic-type evolution
equations.

The Hamiltonian and momentum constraints in vacuum GR are 
\begin{subequations}
\label{eq:ConstraintsGRvac}
\begin{align}
\label{eq:HamGRvac}
\H = &  R - K_{ij} K^{ij} + K^{2}
   = 0
\,,\\
\label{eq:MomGRvac}
\M_{i} = & D^{j} K_{ij} - D_{i} K 
       = 0
\,.
\end{align}
\end{subequations}
To provide initial data $(\gamma_{ij}, K_{ij})|_{t=0}$ describing quasi-circular BH binaries, we employ the Bowen-York
construction~\cite{Bowen:1980yu}.

The time development of the $3$-metric and extrinsic curvature is determined by the evolution equations
\begin{align}
\label{eq:GREvolEqsADM}
\dif_{t} \gamma_{ij} = & - 2 \alpha K_{ij} 
\,,\\
\dif_{t} K_{ij}      = & - D_{i}D_{j} \alpha + \alpha \left[ R_{ij} + K K_{ij} 
- 2 K_{ik} K^{k}{}_{j} \right]
\,,\nonumber
\end{align}
where $\dif_{t} = \p_{t} - \Lie_{\beta}$ and $\Lie_{\beta}$ is the Lie derivative along the shift vector.
To obtain a strongly hyperbolic and, hence, well-posed initial value formulation of Einstein's equations, we employ the
$W$-version of the Baumgarte-Shapiro-Shibata-Nakamura (BSSN) formulation~\cite{Shibata:1995we,Baumgarte:1998te} whose variables
are
\begin{subequations}
\label{eq:WBSSNvars}
\begin{align}
W = & \gamma^{-\frac{1}{6}}
\,,\quad
\tilde{\gamma}_{ij} = W^{2} \gamma_{ij}
\,,\\
K = & \gamma^{ij} K_{ij}
\,,\quad
\tilde{A}_{ij} = W^{2} A_{ij}
\,,\\
\tilde{\Gamma}^{i} = & \tilde{\gamma}^{kl} \tilde{\Gamma}^{i}{}_{kl} 
        = - \p_{k} \tilde{\gamma}^{ik} 
\,,
\end{align}
\end{subequations}
where $\gamma \equiv \det\left(\gamma_{ij}\right)$ is the determinant of the physical $3$-metric, $A_{ij} = K_{ij} -
\frac{1}{3} \gamma_{ij} K$ is the tracefree part of the extrinsic curvature, and the last relation for the conformal
connection function $\tilde{\Gamma}^{i}$ holds because $\tilde{\gamma} =1$ by construction.
The resulting evolution equations are given explicitly, e.g., in Refs.~\cite{ZilhaoWitekCanudaPaper,Clough:2015sqa} and
we complement them with the moving puncture gauge~\cite{Campanelli:2005dd,Baker:2005vv}
\begin{subequations}
\label{eq:MovingPunctureGauge}
\begin{align}
\dif_{t} \alpha = & - 2 \alpha K
\,,\\
\dif_{t} \beta^{i} = & \beta_{\Gamma} \tilde{\Gamma}^{i} - \eta_{\beta} \beta^{i}
\,,
\end{align}
\end{subequations}
where we typically choose 
$\beta_{\Gamma} = 0.75$
and
$\eta_{\beta} = 1$.

\subsection{Gauss--Bonnet invariant in 3+1 form}
Next, we express the Gauss--Bonnet invariant~\eqref{eq:GBInvariant} in terms of spatial quantities.  A particular
convenient reformulation is that in terms of the electric and magnetic parts of the Weyl tensor defined as
\begin{subequations}
\label{eq:SplitWeylEB}
\begin{align}
\label{eq:DefEij}
E_{ij} = & \gamma^{a}{}_{i} \gamma^{b}{}_{j} n^{c} n^{d} W_{acbd}
\,,\\
\label{eq:DefBij}
B_{ij} = & \gamma^{a}{}_{i} \gamma^{b}{}_{j} n^{c} n^{d} \,^{\ast}W_{acbd}
\,,
\end{align}
\end{subequations}
where $\,^{\ast}W_{abcd}$ denotes the dual Weyl tensor.
By construction the electric and magnetic parts of the Weyl tensor are symmetric, tracefree and spatial, {\em i.e.}, $E_{ij} =
E_{(ij)}$, $\gamma^{ij}E_{ij} = 0$, $E_{ab} n^{a} = 0$, and likewise for $B_{ij}$.
Comparing Eqs.~\eqref{eq:SplitWeylEB} to the spacetime decomposition of the Weyl tensor and of the GR field
equations~\eqref{eq:GREvolEqsADM} we find
\begin{subequations}
\begin{align}
B_{ij} = & \epsilon_{(i|}{}^{kl} D_{k} A_{|j)l}
\,,\\
E_{ij} = & R^{\rm{tf}}_{ij} - A_{ik} A^{k}{}_{j} + \frac{1}{3} \left( K A_{ij} 
+ \gamma_{ij} A_{kl} A^{kl} \right)
\,,
\end{align}
\end{subequations}
where $R^{\rm{tf}} = R_{ij} - \frac{1}{3} \gamma_{ij} R$ is the tracefree part of the spatial Ricci tensor.
Assuming a {\textit{Ricci-flat}} background spacetime, {\em i.e.} $^{(4)}R_{ab}=0$, the Gauss--Bonnet
invariant~\eqref{eq:GBInvariant} simplifies to
\begin{align}
\label{eq:RGBinEBvac}
\R_{\rm{GB}} = & 8 \left( E_{ij} E^{ij} - B_{ij} B^{ij} \right)
\,.
\end{align}
This equation, evaluated on the background metric, is the source term in Eq.~\eqref{eq:EoMsEFTapproachEps1}.

\subsection{Scalar-field evolution equations}

To simulate the scalar field's dynamics we rewrite its field equation~\eqref{eq:EoMsEFTapproachEps1} as a set of time
evolution equations.  To this end, we introduce the scalar field momentum
\begin{align}
\label{eq:DefKPhi}
K_{\Phi} = & - \Lie_{n} \Phi
\,,
\end{align}
in analogy to the extrinsic curvature.
Then, the time evolution is determined by
\begin{align}
\label{eq:ScalarEvol}
\dif_{t} \Phi     = & - \alpha K_{\Phi}
\,,\\
\dif_{t} K_{\Phi} = &
        - \alpha \left( D^{i} D_{i} \Phi - K K_{\Phi} + \R_{\rm{GB}} \right)
        - D^{i} \Phi D_{i} \alpha
\,,\nonumber
\end{align}
where the Gauss--Bonnet invariant is calculated from 
Eq.~\eqref{eq:RGBinEBvac}.

\subsection{Scalar-field initial data}~\label{ssec:ScalarInitData}
We will focus on different types of scalar-field initial data that appear most relevant.

\noindent{\bf{Initial data 1 (ID1):}} 
The first set is trivial initial data
\begin{align}
\label{eq:IDSFZero}
\Phi|_{t=0} = & 0
\,,\quad
K_{\Phi}|_{t=0} = 0
\,.
\end{align}
This will allow us to verify the formation of nontrivial scalar hair around rotating BHs or around a BH binary solely
sourced by the spacetime curvature.

\noindent{\bf{Initial data 2 (ID2):}}
To investigate perturbations around isolated BHs we initialize the scalar field and its momenetum as a condensate with a
Gaussian profile centered around $R_{0}$ with width $\sigma$ and amplitude $A$.  Specifically, we set
\begin{align}
\label{eq:IDSFGaussian}
\Phi\vert_{t=0}     = & 0
\,,\\
K_{\Phi}\vert_{t=0} = & A \exp\left[\frac{ (R-R_{0})^{2} }{\sigma^{2}}\right] \Sigma_{lm}(\theta,\varphi)
\,,\nonumber
\end{align}
where
$\Sigma_{lm}$ is a superposition of spherical harmonics, 
typically set to
$\Sigma_{11} = Y_{1-1} - Y_{11}$ or 
$\Sigma_{22} = Y_{22} + Y_{2-2} + Y_{20}$.

\noindent{\bf{Initial data 3 (ID3):}} 
These initial conditions represent data for multiple (hairy) BHs.  For simplicity, we neglect any linear or angular
momenta of the BHs.  Since the scalar field equation~\eqref{eq:EoMsEFTapproachEps1} is linear, we can superpose the
static solution of~\cite{Sotiriou:2013qea}; see~\eqref{eq:SolPhiSmallCouplSmallSpin} with $\chi=0$.  Then, for $N$ BHs,
we have
\begin{align}
\Phi|_{t=0} = & \sum^{N}_{a=1} \Phi_{(a)}
\,,\quad
K_{\Phi}|_{t=0} = 0
\,,
\end{align}
where the field associated to the $(a)$-th BH with ADM mass $m_{(a)}$ and at 
position $R_{(a)}$ is
\begin{align}
\Phi_{(a)} = & \frac{32 R_{(a)} m_{(a)} }{\left( m_{(a)} + 2 R_{(a)} \right)^6 }
               \left[ m^{4}_{(a)} + 12 m^{3}_{(a)} R_{(a)} \right.
\\ & \left.
        + \frac{184}{3} m^{2}_{(a)} R^{2}_{(a)} + 48 m_{(a)} R^{3}_{(a)} + 16 R^{4}_{(a)}
\right]
\,,\nonumber
\end{align}
in the same quasi-isotropic coordinates $(t,R,\theta,\varphi)$ employed to construct the background (GR) initial data.
Finally, we note that our initial data for the scalar fields are strictly valid only when the BHs are all at rest. When
considering a BH binary, we are neglecting the initial velocity of each BH, which could be taken into account by
boosting each of the scalar-field profiles. Neglecting this boost introduces some spurious initial-data effect that is
neglible as the scalar field adjusts to its actual configuration during the evolution.

\subsection{Analysing the data}~\label{ssec:AnalysisTools}
To analyse and interpret the numerical data we extract a number of observables from our simulations each of which we
summarize in the following.

{\noindent{\bf{Waveforms:}}}
To calculate the gravitational radiation produced in our background spacetime we employ the Teukolsky
formalism~\cite{Newman:1961qr,Teukolsky:1973ha}.  In this spinor-inspired approach one defines a null tetrad and a set
of complex scalars that contain information about the radiative degrees of freedom. They are constructed from
contractions of the Weyl tensor with the tetrad vectors.  With an appropriate choice of the tetrad one of these complex
scalars, $\Psi_{4}$, encodes the outgoing gravitational radiation.  For details of the construction see, e.g.,
Refs.~\cite{Alcubierre:2008,Witek:2010qc,Okawa:2014nda}.
In practice, we measure the Newman-Penrose scalar on spheres of fixed extraction radius $\rex$ and decompose it into
multipoles
\begin{align}
\label{eq:DefPsi4lm}
\Psi_{4,lm}(t,\rex) = & \int \dif \Omega\, \Psi_{4}(t,\rex,\theta,\varphi) 
_{-2}Y^{\ast}_{lm}(\theta,\varphi)
\,,
\end{align}
where $_{-2}Y_{lm}$ are $s=-2$ spin-weighted spherical harmonics.

In a similar fashion we extract scalar radiation: we interpolate the scalar field $\Phi$ onto a sphere of radius $\rex$
and perform a multipole decomposition using spherical harmonics $Y_{lm}$.  In particular, we measure
\begin{align}
\label{eq:DefPhilm}
\Phi_{lm} (t,\rex) = & \int\dif\Omega\, \Phi(t,\rex,\theta,\varphi) 
Y^{\ast}_{lm}(\theta,\varphi)
\,.
\end{align}

{\noindent{\bf{Energy and momentum fluxes:}}}
In addition to the waveforms, the energy and momentum fluxes provide crucial insight into the phenomenology of the
system and allow us to estimate the order-of-magnitude of metric deformations and radiation at second order without
actually evolving it.
First, let us recap the energy and momentum fluxes of the gravitational radiation in
GR~\cite{Wald:1984rg,Alcubierre:2008}.  These are fluxes present in the background spacetime and given by
\begin{subequations}
\label{eq:FluxesGWGR}
\begin{align}
\label{eq:EnergyFluxGWGR}
\frac{\dif E^{\rm{GW}}}{\dif t} = &
  \lim_{R\rightarrow\infty} \frac{R^{2}}{16\pi} 
\int\dif\Omega\,\left|\int_{-\infty}^{t} \Psi_{4}\dif\tilde{t}\,\right|^{2}
\,,\\
\label{eq:MomentumFluxGWGR}
\frac{\dif P^{\rm{GW}}_{i}}{\dif t} = &
- \lim_{R\rightarrow\infty} \frac{R^{2}}{16\pi} 
\int\dif\Omega\,\ell_{i}\,\left|\int_{-\infty}^{t} 
\Psi_{4}\dif\tilde{t}\,\right|^{2}
\,,
\end{align}
\end{subequations}
where 
$\ell = -\left( \sin\theta\cos\varphi, \sin\theta\sin\varphi,\cos\theta\right)$.

Furthermore, we consider the energy and momentum fluxes of the scalar field, which are associated to the scalar
stress-energy tensor $T_{ab}$ and to the Gauss--Bonnet correction ${\cal G}_{ab}$.  For a generic energy-momentum tensor
${\cal T}_{ab}$ 
they can be defined as~\cite{Wald:1984rg,Alcubierre:2008,Teukolsky:1973ha}
\begin{subequations}
\label{eq:FluxesGeneral}
\begin{align}
\frac{\dif E}{\dif t} = & \lim_{R\rightarrow\infty} R^{2} \int\, \dif\Omega\, 
{\cal T}_{0R}
                      = - \lim_{R\rightarrow\infty} R^{2} \int\, \dif\Omega\, 
j_{R}
\,,\\
\frac{\dif P_{i}}{\dif t} = & \lim_{R\rightarrow\infty} R^{2} \int\, 
\dif\Omega\, {\cal T}_{iR}
                      =   \lim_{R\rightarrow\infty} R^{2} \int\, \dif\Omega\, 
S_{iR}
\,,
\end{align}
\end{subequations}

\noindent where $j_{R}$ and $S_{iR}$ are the radial energy-momentum flux and stress tensor, generically computed from
\begin{align}
\label{eq:DefTmn3p1}
j_{i} = & - \gamma^{a}{}_{i} n^{b} {\cal T}_{ab}
\,,\quad
S_{ij} = \gamma^{a}{}_{i} \gamma^{b}{}_{j} {\cal T}_{ab}
\,.
\end{align}
Since our code is implemented explicitly in Cartesian coordinates, we transform these quantities to spherical
coordinates
\begin{align}
\label{eq:TrafoTmnSph2Cart}
j_{R} = & \Lambda^{k}{}_{R} j_{k}
\,,\quad
S_{iR} = \Lambda^{k}{}_{R} S_{ik}
\,,
\end{align}
where $\Lambda^{i}{}_{a} = \frac{\dif X^{i}}{\dif X^{a}} $ is defined by transformation~\eqref{eq:TrafoSphVsCart}.

Since $\Phi={\cal O}(\epsilon)$, the leading-order components of the scalar energy-momentum tensor and of the
Gauss--Bonnet correction [{\em i.e.}, the source terms of modified Einstein's equations~\eqref{eq:ExpansionEps2Phi0Metric}]
are ${\cal O}(\epsilon^2)$. 
%
%
Using the effective stress-energy tensor defined in Eq.~\eqref{eq:ExpansionEps2TmnEff} we can write the
(leading order) energy flux carried by the scalar field as
fluxes as
\begin{align}
\label{eq:DefFluxes2ndOrderSum}
\frac{\dif E^{(2)}}{\dif t} = & \frac{\dif E^{(\Phi)}}{\dif t} - 16M^2 \frac{\dif 
E^{\rm{GB}}}{\dif t}
\,,
\end{align}
and likewise for the momentum flux.  Here the energy and momentum fluxes for the scalar field
associated to the canonical energy-momentum tensor, indicated by ${}^{(\Phi)}$, 
and those associated to the
Gauss--Bonnet correction, indicated by ${}^{\rm{GB}}$, are obtained from Eqs.~\eqref{eq:FluxesGeneral} using the
relevant flux densities and spatial stress tensors.  Using~\eqref{eq:DefTmn3p1} and replacing ${\cal T}_{ab}$ with
$T^{(2)}_{ab}$ (of the scalar) defined in Eq.~\eqref{eq:ExpansionTmnSFEps2} we find~\footnote{Recall that we suppress
  superscripts indicating the order and bear in mind that $(g_{ab},\Phi)\equiv(g^{\rm{GR}}_{ab},\PhiI)$.}
\begin{subequations}
\label{eq:jiSijPhi}
\begin{align}
j^{(\Phi)}_{i}  = & K_{\Phi} D_{i} \Phi
\,,\\
S^{(\Phi)}_{ij} = &  D_{i}\Phi D_{j}\Phi + \frac{1}{2}\gamma_{ij} \left( 
K^{2}_{\Phi} - D^{k}\Phi D_{k}\Phi \right)
\,.
\end{align}
\end{subequations}
If, instead, we replace ${\cal T}_{ab}$ with $\G^{(1)}_{ab}$ defined in Eq.~\eqref{eq:ExpansionTmnGBEps1} and
insert~\eqref{eq:FPhiExplicit} we get
\begin{subequations}
\label{eq:jiSijGB}
\begin{align}
j^{\rm{GB}}_{i} = & E^{k}{}_{i} \left( D_{k} K_{\Phi} - K^{l}{}_{k} D_{l}\Phi \right)
\nonumber \\ & 
        + \epsilon_{i}{}^{jk} B^{l}{}_{j} \left( D_{k}D_{l}\Phi - K_{\Phi} K_{kl} \right)
\,,\\
S^{\rm{GB}}_{ij} = & 2 \left(\bar{S}^{\rm{GB}}_{ij} - \frac{1}{2}\gamma_{ij} 
\gamma^{kl}  \bar{S}^{\rm{GB}}_{kl} \right)
\,,
\end{align}
\end{subequations}
where  
\begin{align}
\bar{S}^{\rm{GB}}_{ij} = &
  E^{l}{}_{(i|} \left( K_{\Phi} K_{|j)l} - D_{|j)}D_{l}\Phi \right)
+ E_{ij} \left( D^{l} D_{l} \Phi - K K_{\Phi} \right)
\nonumber \\ &
+ \epsilon_{(i}{}^{kl} B_{j)k} \left(D_{l}K_{\Phi} - K^{m}{}_{l} D_{m}\Phi 
\right)
\,.
\end{align}

In practice, we take the following steps:
\begin{enumerate*}[label={(\roman*)}]
\item we compute $j_{i}$ and $S_{ij}$ during the numerical evolution from Eqs.~\eqref{eq:jiSijPhi} and~\eqref{eq:jiSijGB};
\item since the numerical code is in Cartesian coordinates we perform a coordinate
  transformation~\eqref{eq:TrafoTmnSph2Cart};
\item we use the radial fluxes to calculate the energy and momentum fluxes through Eqs.~\eqref{eq:FluxesGeneral}.
\end{enumerate*}

We remark that at ${\cal O}(\epsilon^2)$ the energy flux contains a
contribution from gravitational radiation coming from ``mixed'' terms
$\sim \langle{\dddot h}^{(0)}_{ab}{\dddot h}^{(2)}_{ab}\rangle$,
where $h^{(0)}_{ab}$ denotes the background contribution. 
We do not explicitly compute $h^{(2)}_{ab}$, since it would 
require to numerically evolve Einstein's equations at ${\cal O}(\epsilon^2)$.
However, 
it is reasonable to assume that this contribution is at
most comparable to the energy flux carried by the scalar field. In the inspiral phase, it is expected to be sub-dominant
with respect to the scalar field flux, since it is of higher post-Newtonian~(PN)
order~\cite{Okounkova:2017yby}. Therefore, Eq.~\eqref{eq:DefFluxes2ndOrderSum} provides a reliable estimate of the
energy flux at ${\cal O}(\epsilon^2)$.

\section{Results}~\label{sec:NumericalResults}

\subsection{Code description}
To simulate BHs in sGB gravity we implemented the field equations~\eqref{eq:ConstraintsGRvac},~\eqref{eq:GREvolEqsADM}
and~\eqref{eq:ScalarEvol} in \canuda~\cite{ZilhaoWitekCanudaPaper,ZilhaoWitekCanudaRepository}.  \canuda~\footnote{The
  name is inspired by the ``Cemetery of Forgotten Books'' series by C.~R.~Zaf{\'{o}}n that, in turn, was inspired by a
  historic library of that name in Barcelona.}  is a novel numerical relativity library that is compatible with the
open-source \ETK~\cite{Loffler:2011ay,Zilhao:2013hia,EinsteinToolkit:web}, and capable to simulate BHs in extensions of
GR; see, e.g.~\cite{Okawa:2014nda,Zilhao:2015tya,ZilhaoWitekCanudaPaper}.
The \ETK\, itself is a community code originally designed to solve the two-body problem in GR.  It is based on the
{\textsc{Cactus}} computational toolkit~\cite{Goodale:2002a,Cactuscode:web} and the {\textsc{Carpet}} boxes-in-boxes
adaptive mesh refinement package~\cite{CarpetCode:web}.

In practice, we evolve the background BH spacetime and the scalar field simultaneously, as described in
Sec.~\ref{sec:EvolFormulation}.
We set up BH initial data using either the {\textsc{TwoPunctures}} spectral code for binaries~\cite{Ansorg:2004ds} or
fix the background as a single rotating BH as described in Sec.~\ref{sssec:KerrInQIcoords}.
To evolve BH binaries we typically employ \canuda--\lean\, (an upgraded version of~\cite{Sperhake:2006cy}), although our
implementation can be combined with other \ETK\, evolution thorns such as {\textsc{McLachlan}}~\cite{Brown:2008sb}.
We extract apparent or isolated horizon properties of the BHs with
{\textsc{AHFinderDirect}}~\cite{Thornburg:2003sf,Thornburg:1995cp} and
{\textsc{Quasilocalmeasures}}~\cite{Dreyer:2002mx}.
For the wave extraction we typically use our own implementation of the Weyl scalars or the built-in \ETK\, version
thereof.

The core thorns~\footnote{``Thorns'' refer to code modules in \ETK\, speak.}  developed for the purpose of the present
paper are already publicly available~\cite{ZilhaoWitekCanudaRepository}, and consist of an initial data thorn
implementing the prescription of Sec.~\ref{ssec:ScalarInitData} and an evolution thorn implementing
Eqs.~\eqref{eq:ScalarEvol} in the $W$-version of the BSSN formulation and analysis capabilities as described in
Sec.~\ref{ssec:AnalysisTools}.

\subsection{Hair formation around rotating black holes}
Following up on previous studies~\cite{Benkel:2016rlz,Benkel:2016kcq}
we explore the formation of scalar hair around rotating BHs.
Stationary solutions have been obtained analytically or generated numerically (see
Appendix~\ref{ssec:BHsolsAna}). Their key feature of the most general family of such solutions is that it is
generically singular on the horizon. Imposing that the solution is regular across the horizon selects a sub-family of
solutions that is uniquely characterised by the mass and the spin of the 
black hole. The corresponding scalar configuration is unique, {\em i.e.}~there are no independent scalar
charges~\cite{Kanti:1995vq,Sotiriou:2014pfa}. This kind of nontrivial scalar configuration around BHs is also known as
scalar hair of the second kind. Though it seems natural to discard the singular configurations (as taking the
stationary limit of a PDE can 
lead to dynamically spurious singular solution), it is crucial to demonstrate
explicitly that the regular solutions are indeed the endpoints of dynamical processes, such as collapse and mergers.

{\noindent{\textbf{Setup:}}} To explore this process we performed a set of time domain simulations for different initial
configurations of the scalar field around rotating BHs with total mass $M=1$ and various dimensionless spins,
$\chi\in[0,0.99]$.  We initialized the scalar field either as vanishing or as a Gaussian composed of the dipole or
quadrupole mode located at $R_{0}=10M$ with width $\sigma=1M$ and amplitude $A=1M$ denoted, respectively, as Initial
data $1$ or $2$ in Sec.~\ref{ssec:ScalarInitData}.
The different configurations are summarized in Table~\ref{tab:KerrSetup}.

Our numerical domain with outer boundaries at $256M$ consisted of nine refinement levels centered around the BH.  On the
outermost level we typically set the grid-spacing to $dx=2.0M$.
\begin{table}[htpb!]
\caption{\label{tab:KerrSetup} Parameters for runs modelling the formation of scalar hair around rotating BHs with
  dimensionless spin $\chi$.  We initialize the scalar field as vanishing
  (ID1 in Sec.~\ref{ssec:ScalarInitData}) or as Gaussian~\eqref{eq:IDSFGaussian} 
  (ID2 in Sec.~\ref{ssec:ScalarInitData}) containing dipole ($\Sigma_{11}$) or quadrupole ($\Sigma_{22}$) modes.  }
\begin{tabular}{|l|c|c|}
\hline
Run       & $\chi$ & scalar initial data                        \\
\hline
\hline
Kerr\_0   & $0.0$  & ID1; ID2: $\Sigma_{11}$                    \\
Kerr\_02  & $0.2$  & ID1                                        \\
Kerr\_05  & $0.5$  & ID1; ID2: $\Sigma_{11}$                    \\
Kerr\_07  & $0.7$  & ID1; ID2: $\Sigma_{11}$; $\Sigma_{22}$     \\
Kerr\_09  & $0.9$  & ID1; ID2: $\Sigma_{11}$                    \\
Kerr\_099 & $0.99$ & ID1; ID2: $\Sigma_{11}$; $\Sigma_{22}$     \\
\hline
\end{tabular}
\end{table}

\begin{figure}[htpb!]
\begin{center}
\includegraphics[width=0.5\textwidth,clip]{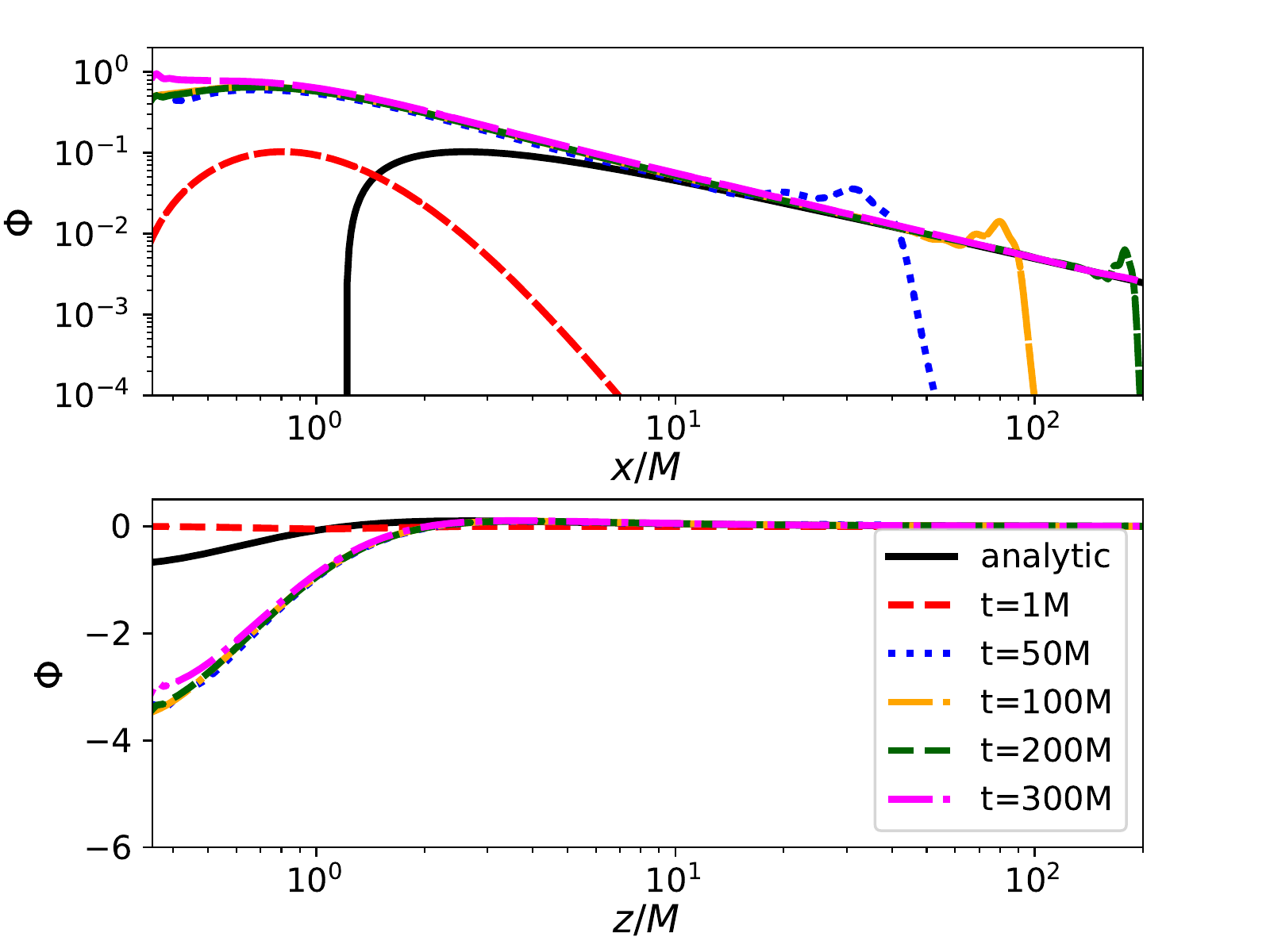}
\caption{\label{fig:Kerr_PhiProfile_a099_VarTime} (Color online)  Profile of an initially vanishing scalar field excited
  by a BH with dimensionless spin $\chi=0.99$ at different instances throughout its evolution.  We present it along the
  equatorial plane (top panel) and perpendicular to it (bottom panel) At late times and large distances, $R/M\gg1$, we
  approach the analytic solution~\eqref{eq:SolPhiArbSpinLargeR} indicated by the solid black curve. 
}
\end{center}
\end{figure}

{\noindent{\textbf{Scalar evolution:}}} In Fig.~\ref{fig:Kerr_PhiProfile_a099_VarTime} we illustrate the formation of
scalar hair exemplarily around a highly rotating BH with dimensionless spin $\chi=0.99$.  Specifically, we present the
scalar field profile $\Phi$ along the x-axis (equatorial plane) and z-axis (direction aligned with BH spin) at different
instances in time.
To compare the numerically obtained field after an evolution time of $t=300M$ to the analytic solution in a large radius
expansion (see Eq.~\eqref{eq:SolPhiArbSpinLargeR}), we transform the latter to quasi-isotropic coordinates that coincide
with the numerical ones for $R\gtrsim5$; see Fig.~A1 of~\cite{Benkel:2016rlz}.
We find excellent agreement between the numerical and analytic solution at large distances.

\begin{figure}[htpb!]
\begin{center}
\includegraphics[width=0.5\textwidth,clip]{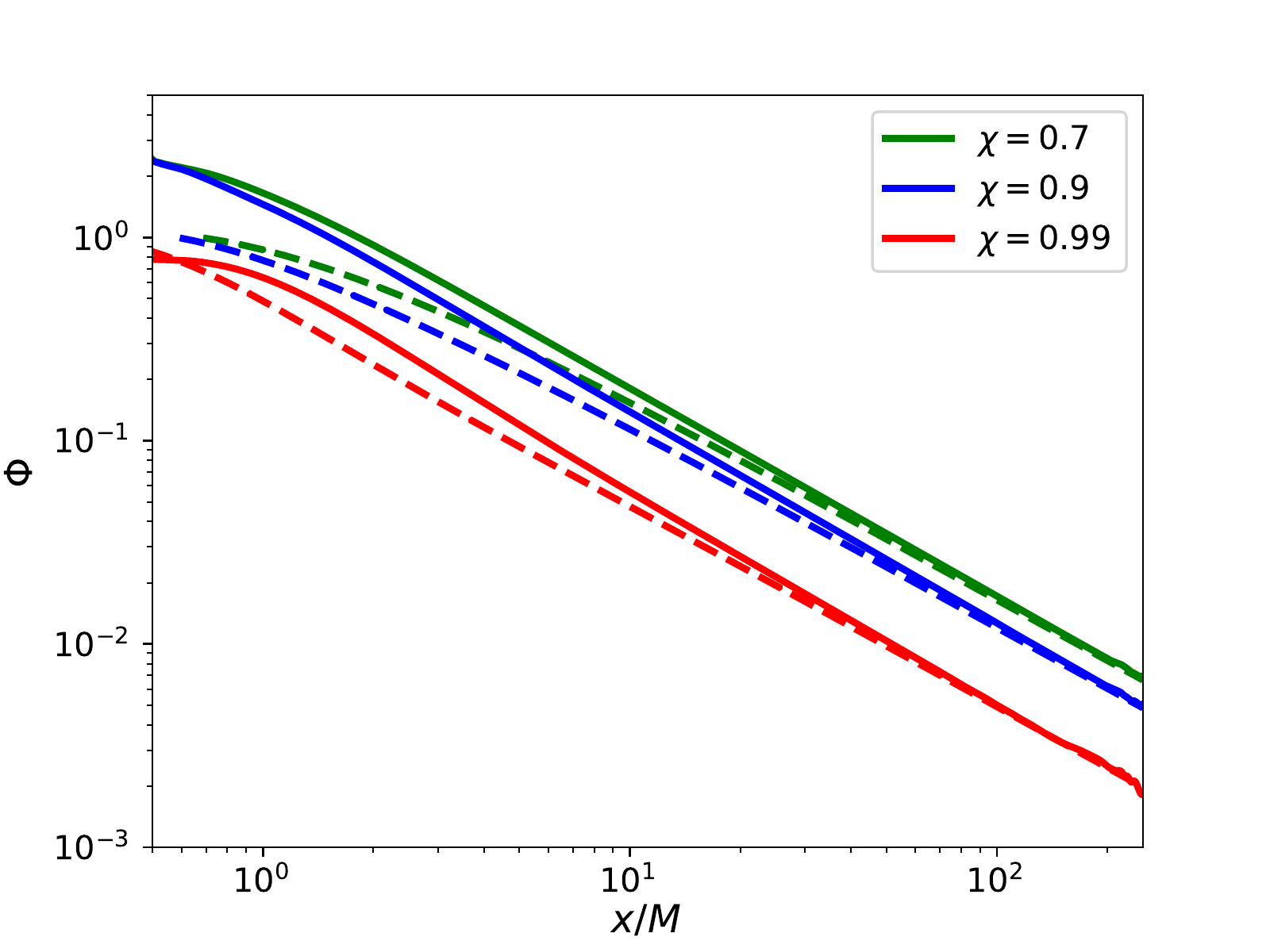}
\caption{\label{fig:Kerr_PhiProfile_NumVsAna_t300_VarSpin} (Color online) Comparison between the numerically evolved
  scalar field (solid curves) at late times $t=300M$ and the analytic solutions~\eqref{eq:SolPhiArbSpinLargeR} (dashed
  curves) for different values of the BH spin.  }
\end{center}
\end{figure}

Near the horizon, the analytical solution is less accurate mostly in the polar direction, whereas on the equatorial
plane it provides a good approximation to the exact numerical result also close to the horizon.
We further quantify that statement in Fig.~\ref{fig:Kerr_PhiProfile_NumVsAna_t300_VarSpin} where we benchmark the
numerical scalar field solution after $t=300M$ against the analytic one, Eq.~\eqref{eq:SolPhiArbSpinLargeR}.  We focus
on highly spinning BH backgrounds and find excellent agreement for sufficiently large distance where the analytic
approximation is valid.


\begin{figure*}[htpb!]
\begin{center}
\includegraphics[width=0.48\textwidth,clip]{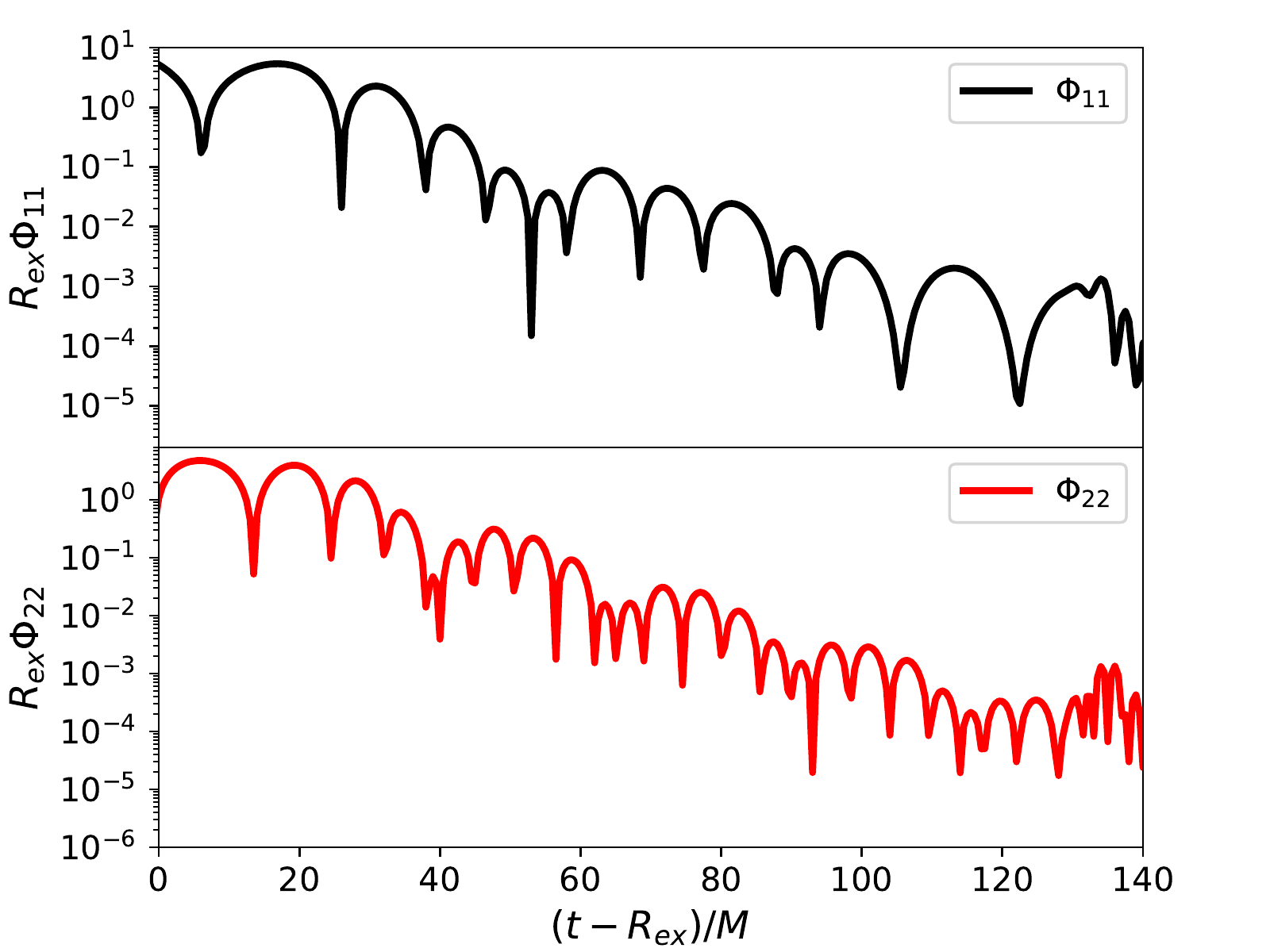} 
\includegraphics[width=0.48\textwidth,clip]{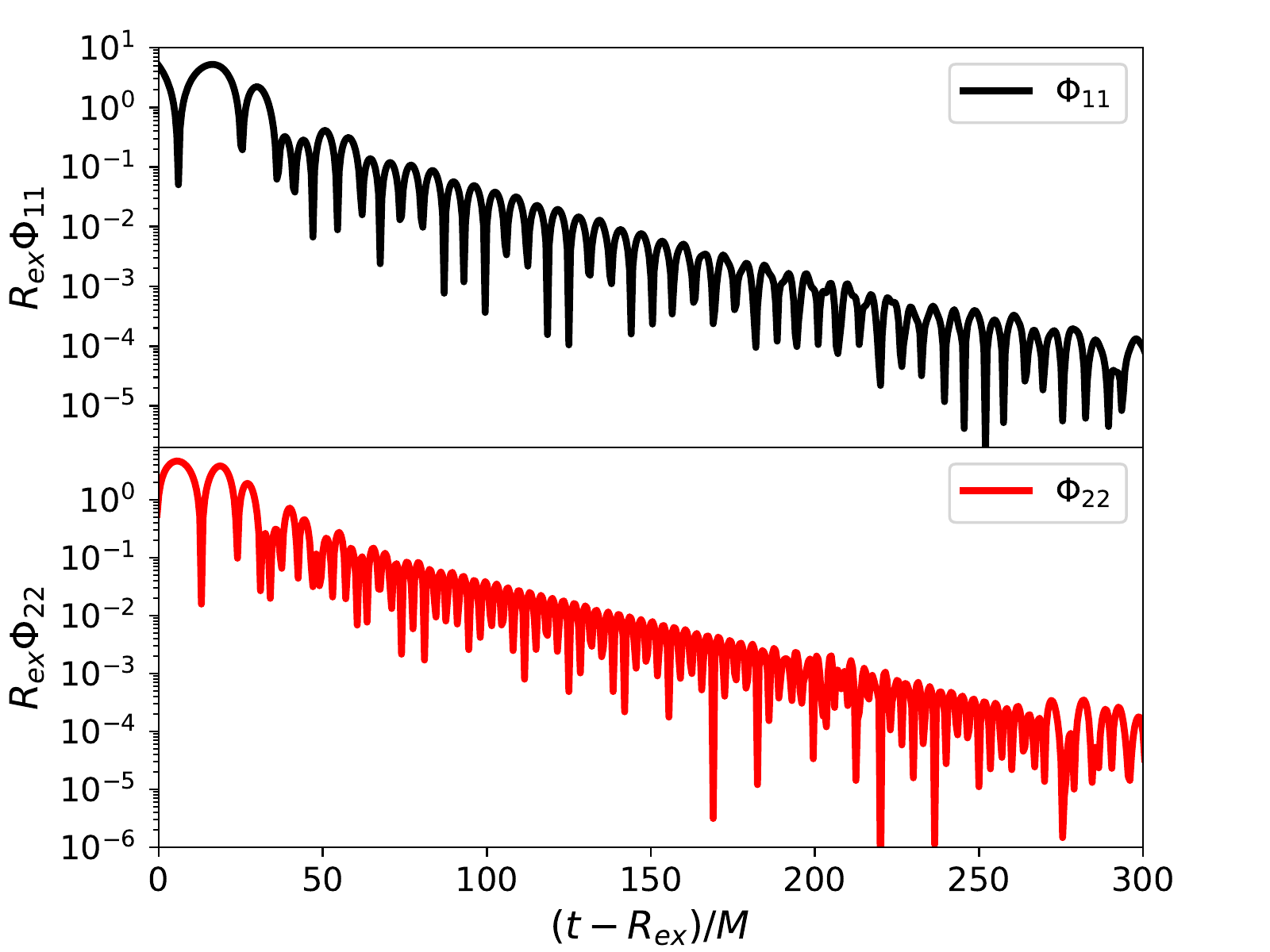}
\caption{\label{fig:Kerr_Waveforms_a070090_rex60} (Color online) Scalar field $l=m=1$ (top) and $l=m=2$ (bottom)
  multipoles evolved around a BH with dimensionless spin $\chi=0.7$ (left panel) and $\chi=0.99$ (right panel).  We
  rescale it by the extraction radius $\rex=60M$, and shift it in time.
  For $\chi=0.7$, the $l=m=2$ mode exhibits a significant modulation indicating a comparable excitation (and
  superposition) of the scalar-led and gravitational-led modes.
  For $\chi=0.99$ both multipoles are consistent with the (free) scalar quasi-normal ringdown.  }
\end{center}
\end{figure*}

{\noindent{\textbf{Waveforms:}}} As the system settles down to a hairy BH in sGB gravity, the scalar field sheds away
all non axisymmetric modes.  This is illustrated in Fig.~\ref{fig:Kerr_Waveforms_a070090_rex60}, where we show the
$l=m=1,2$ modes of the scalar, measured at $\rex = 60M$, around a BH with dimensionless spin $\chi=0.7$ and $\chi=0.99$.
The latter case exhibits a decay pattern that is consistent with the quasi-normal ringdown of a free scalar around a
Kerr BH~\cite{Berti:2009kk}, {\em i.e.}, it is determined by the dominant (scalar) QNM frequency.
In contrast, the $\chi=0.7$ case shows a more complex, modulated ringdown that indicates the presence of multiple modes
with comparable excitation frequency and amplitude.
Inspecting Eq.~\eqref{eq:EoMsEFTapproachEps1} this is not surprising; the scalar field is sourced by the Gauss--Bonnet
invariant.  Hence, we expect the superposition of two types of modes\footnote{Similar behaviour has been found in the
  case of BHs in dynamical Chern-Simons~\cite{Molina:2010fb} and nonrotating BHs in Gauss--Bonnet
  gravity~\cite{Blazquez-Salcedo:2016enn}.  The rotating case studied here, however, is largely unexplored and might
  contain additional effects.}:
\begin{enumerate*}[label={(\roman*)}]
\item a scalar mode corresponding to the decay of a free scalar field around a Kerr BH;
\item a mode driven by the background curvature.
\end{enumerate*}
We refer to the former as ``scalar-led'' and the latter as ``gravitational-led'' modes.
The signal's specific morphology then depends on the amplitude with which each of those modes are excited.
To identify the composition of the ringdown signal we perform a two-mode fit
\begin{align}
\label{eq:KerrTwoModeFit}
\Phi = & A \left( e^{\omega_{\rm{1I}} t} \cos\left[\omega_{\rm{1R}} t + 
\delta\phi 
\right]              + \delta A_{\rm{2}} e^{\omega_{\rm{2I}}t} 
\cos\left[\omega_{\rm{2R}}t \right]
\right)
\,
\end{align}
where we suppress multipole indices $(lm)$, we indicate the dominant and subdominant mode by numeral subscripts, their
real and imaginary parts by subscripts ``R'' or ``I'', and the relative amplitude between the modes is $\delta
A_{\rm{2}} = A_{\rm{2}} / A$.
We summarize the results in Table~\ref{tab:KerrQNMmodes}.  In particular, we provide estimates of the dominant ringdown
frequency $\omega_{1}$, the real part of the secondary mode $\omega_{2}$, and their relative amplitude.  In practice, we
cannot accurately estimate the decay rate $\omega_{\rm{2I}}$ of the secondary mode.
Additionally, we compare our time-domain estimates to frequency domain calculations of scalar QNMs~\cite{Berti:2009kk}
and find agreement within $\lesssim6\%$.
If the $l=2$ multipole is present in the scalar initial data and, hence, in the ringdown signal, the secondary,
gravitational-led mode's frequency coincides with that of a $l=m=2$ gravitational perturbation.
Although there is no gravitational analogue of the dipole mode, we find good agreement with a back-of-the-envelope
estimate $\omega^{\rm{G}}_{11} \sim \omega^{\rm{G}}_{22}/2$.
Both findings support our expectation of the presence of a gravitational-led mode in the scalar field emission.

There are, however, some caveats in performing these fits: the early response is followed by a (in most cases)
relatively short ringdown signal that transitions to the late-time tail (not shown in the plots).  Combined with the
uncertainty regarding the end of the direct response and the starting point of the actual ringdown -- a hindrance that
is not even completely resolved within GR~\cite{Bhagwat:2017tkm,Brito:2018rfr} -- this results in the uncertainties
quoted above.

\begin{table}[htpb!]
\caption{\label{tab:KerrQNMmodes} Ringdown frequencies of decaying scalar field. We denote the dominant and the (real
  part of the) first subdominant mode, estimated from the time domain data, as $\omega_{\rm{1,lm}}$ and
  $\omega_{\rm{2R,lm}}$, respectively.  Their estimated relative amplitude is $\delta A_{\rm{2}} = A_{\rm{2}}/A$.  
  A dash indicates that the ringdown is dominated by a single mode and we could not accurately estimate a secondary one.
  In the last column we provide the scalar QNM frequency $\omega^{\rm{f,S}}_{lm}$ of a Kerr BH calculated in the
  frequency domain~\cite{Berti:2009kk}.  Where a secondary mode is present it agrees well with the corresponding
  gravitational mode (in case of $l=m=2$) or with $\sim\omega^{\rm{f,S}}_{22}/2$ (in case of $l=m=1$). As a reference,
  the $l=m=2$ gravitational QNM of a Kerr BH with $\chi =0.7$ is $\omega_{22}^{\rm f,G}\approx 0.5326$.  }
\begin{tabular}{|l|c|c|c|c|c|}
\hline
Run       & (lm) & $\omega_{\rm{1,lm}}$ & $\omega_{\rm{2R,lm}}$ & $\delta 
A_{\rm{2}}$ & $\omega^{\rm{f,S}}_{lm}$ \\
\hline
Kerr\_0   & (11) & $0.283  -\imath0.097$    & --                        & --    
                 & $0.2929 -\imath0.0977$   \\
Kerr\_05  & (11) & $0.324  -\imath0.091$    & --                        & --    
                 & $0.3448 -\imath0.0944$   \\
Kerr\_07  & (11) & $0.379  -\imath0.086$    & $0.266$                   & $0.6$ 
                 & $0.3792 -\imath0.0888$   \\
          & (22) & $0.651  -\imath0.089$    & $0.535$                   & $0.4$ 
                 & $0.6561 -\imath0.0876$   \\
Kerr\_09  & (11) & $0.437  -\imath0.071$    & $0.361$                   & $0.2$ 
                 & $0.4372 -\imath0.0718$   \\
Kerr\_099 & (11) & $0.498  -\imath0.035$    & --                        & --    
                 & $0.4934 -\imath0.0367$   \\
          & (22) & $0.930  -\imath0.033$    & --                        & --    
                 & $0.9280 -\imath0.0311$   \\
\hline
\end{tabular}
\end{table}
We expect that the occurrence of a secondary, gravitational-led mode to be a generic feature for all spins, but in some
cases (indicated by a dash in Table~\ref{tab:KerrQNMmodes}) with excitation factors that could be too small to be
extracted from our fit.
This seems to be the case for small spin and near extremality. Interestingly, the maximum excitation factor of the
secondary mode seems to occur for the case $\chi\approx0.7$, which is also phenomenologically relevant since it is
approximately the final spin of a BH remnant from the coalescence of two slowly-spinning BHs. In this case our analysis
predicts the emission of scalar radiation at two dominant frequencies, the scalar and the gravitational fundamental QNMs
of the corresponding Kerr BH.
We also expect that this effect occurs at higher order. In particular, the corrections to the gravitational ringdown
(not computed here since they are of ${\cal O}(\epsilon^2)$)
should contain modes of the ${\cal O}(\epsilon)$ scalar
field, in particular the $l=m=1$ and $l=m=2$ scalar QNMs a Kerr BH. Computing the excitation factors of the latter
requires to solve the field equations at ${\cal O}(\epsilon^2)$ and is left for future work.

\subsection{Black hole binaries}
{\textbf{Setup:}} We focus on nonrotating BHs because they are the simplest BH solutions in sGB gravity that develop
scalar hair.  The BHs initial separation for the results presented below is $d=10M$, though we have considered other
initial distances as well. The system's total mass $M=m_{1}+m_{2}=1$ and the mass ratio varies between
$q=m_{1}/m_{2}=1,1/2,1/4$.  We summarize details of the inital BHs' parameters and the final state in
Table~\ref{tab:BBHSetup}.
This includes the dimensionless spin $\chi_{f}$ of the final BH computed from~\cite{Witek:2010qc}
\begin{align}
\chi_{f} = & \sqrt{ 1 - \left(\frac{2\pi A_{\rm{AH}}}{C^{2}_{\rm{e}} } - 1 \right)^{2} }
\,,
\end{align}
where $A_{\rm{AH}}$ and $C_{\rm{e}}$ denote the area and equatorial circumference of the apparent horizon, its
Christodoulou mass
\begin{align}
M^{2}_{\rm{f}} =& M^{2}_{\rm{irr}} + \frac{J^{2}}{4 M^{2}_{\rm{irr}}}
\,,
\end{align}
where $M_{\rm{irr}}=A_{\rm{AH}}/16\pi$ is the irreducible mass, and the energy $E_{\rm{GW}}/M$ radiated in GWs.
These results are in good agreement with~\cite{Berti:2007fi}
studying unequal mass BH collisions in GR.

\begin{table*}[htpb!]
\caption{\label{tab:BBHSetup} Parameters for nonspinning BH binaries with total mass $M=m_{1}+m_{2}=1$, mass ratio
  $q=m_{1}/m_{2}$ and symmetric mass ratio $\eta = m_{1} m_{2}/M^{2}$.  $m_{i}$ denote the physical BH masses.  Their
  initial separation is $d = x_{1} - x_{2} = 10M$ (situated along the x-axis).  We denote their orbital and tangential
  momenta $P_{r} = P_{2x}=-P_{1x} $ and $P_{\perp} = P_{1y} = -P_{2y}$.
We also denote the dimensionless spin $\chi_{f}$ and mass $M_{\rm{f}}$ of the remnant BH, and the energy
$E^{\rm{GW}}_{\infty}/M$ radiated in form of GWs and the second-order contribution $\epsilon^2E^{\rm{(2)}}_{\infty}/M$
extrapolated to $R\rightarrow\infty$.  }
\begin{tabular}{|l|c|c|c|c|c|c|c|c|c|c|}
\hline
Run     & $q$   & $\eta$& $(x_{1}, x_{2})$  & $-P_{1x}\times10^{4}$ & $P_{1y}$  & $\chi_{f}$ & $M_{\rm{f}}$ 
        & $10^{2} E^{\rm{GW}}_{\infty}/M$ 
& $\epsilon^2 E^{\rm{(2)}}_{\infty}/M$  \\
\hline
BBH\_q1 & $1$   & $1/4$ & $(5,-5)$          & $9.79$                & 
$0.0962578$ & $0.69$ & $0.9596$     & $3.7$ & $0.29\epsilon^2$
\\
\hline
BBH\_q2 & $1/2$ & $2/9$ & $(6.6594,-3.3406)$& $3.78$                & 
$0.0856592$ & $0.62$ & $0.9662$     & $2.9$ & $1.32\epsilon^2$
\\
\hline  
BBH\_q4 & $1/4$ & $4/25$& $(7.9903,-2.0097)$& $3.63$                & 
$0.0618307$ & $0.47$ & $0.9792$     & $1.5$ & $5.85\epsilon^2$
\\
\hline
\end{tabular}
\end{table*}

\begin{figure}[htpb!]
\includegraphics[width=0.5\textwidth,clip]{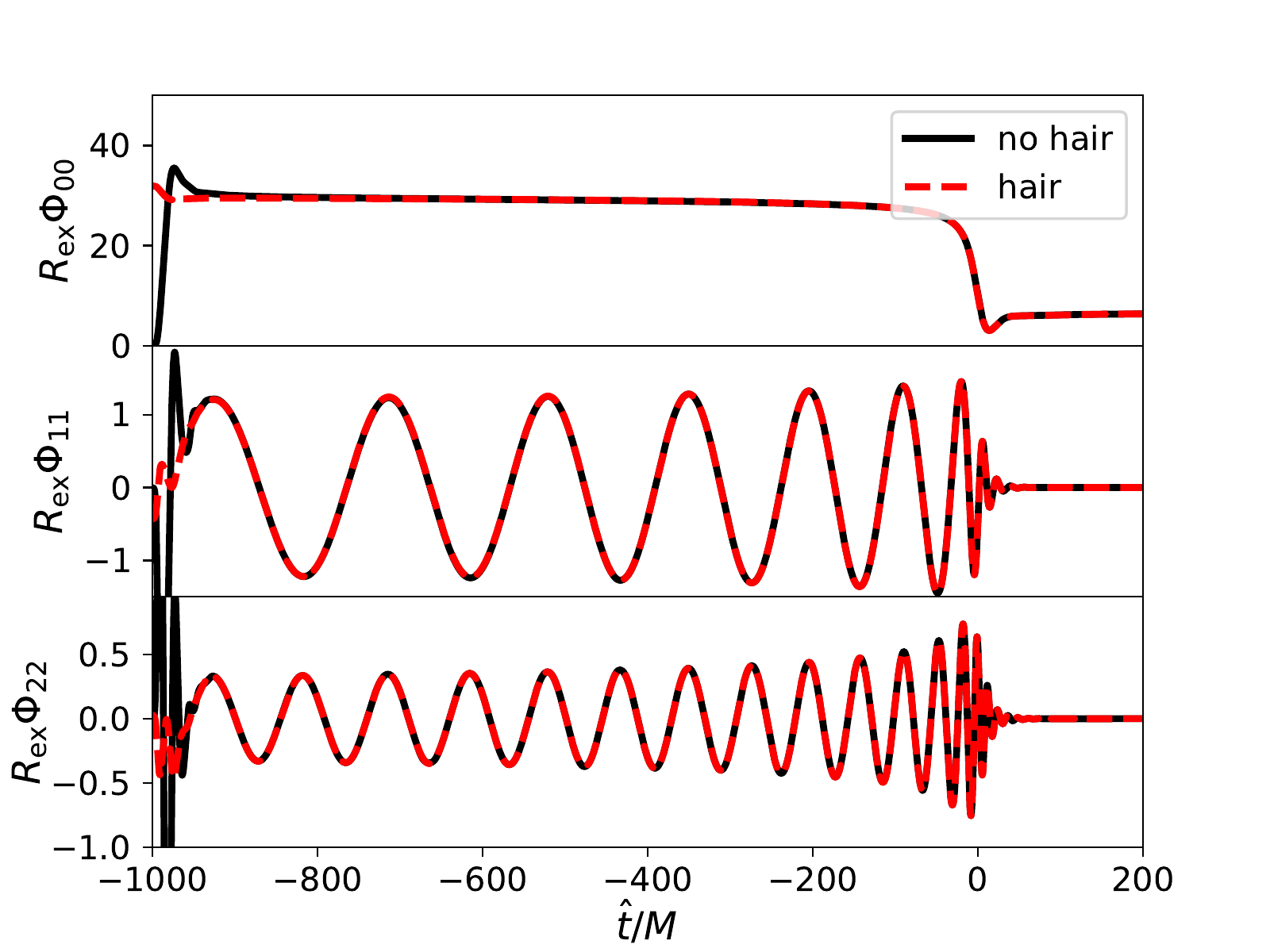}
\caption{\label{fig:BBHPhi_q12_HairVsSF0}
(Color online) Scalar waveforms sourced by a BH binary with mass ratio $q=1/2$. We compare their evolution starting from
  different initial data, namely an initially zero scalar field (solid black curves) and scalar hair corresponding to
  the solution around each of the BHs (red dashed curves); cf. ID3 in Sec.~\ref{ssec:ScalarInitData}. 
We present the scalar's $l=m=0$ (top), $l=m=1$ (middle) and $l=m=2$ (bottom) modes, rescaled by the extraction radius
$R_{\rm{ex}}=100M$ and shifted in time so that $\hat{t}=0$ indicates the time of merger.
After the initial transient during which the scalar field adjusts itself to the hairy BH solutions the evolution of both
cases coincides.}
\end{figure}

We considered scalar initial data ID1 and ID3 
of Sec.~\ref{ssec:ScalarInitData}, {\em i.e.} either starting with an initially vanishing scalar or superposing two
hairy solutions.  Except for an early transition or built-up period both setups yield equivalent results as we
illustrate in Fig.~\ref{fig:BBHPhi_q12_HairVsSF0} for the case of BHs with a mass ratio $q=1/2$.  Therefore, in the
following we present results only for initially hairy BHs.

\begin{figure}[htpb!]
\includegraphics[width=0.5\textwidth,clip]{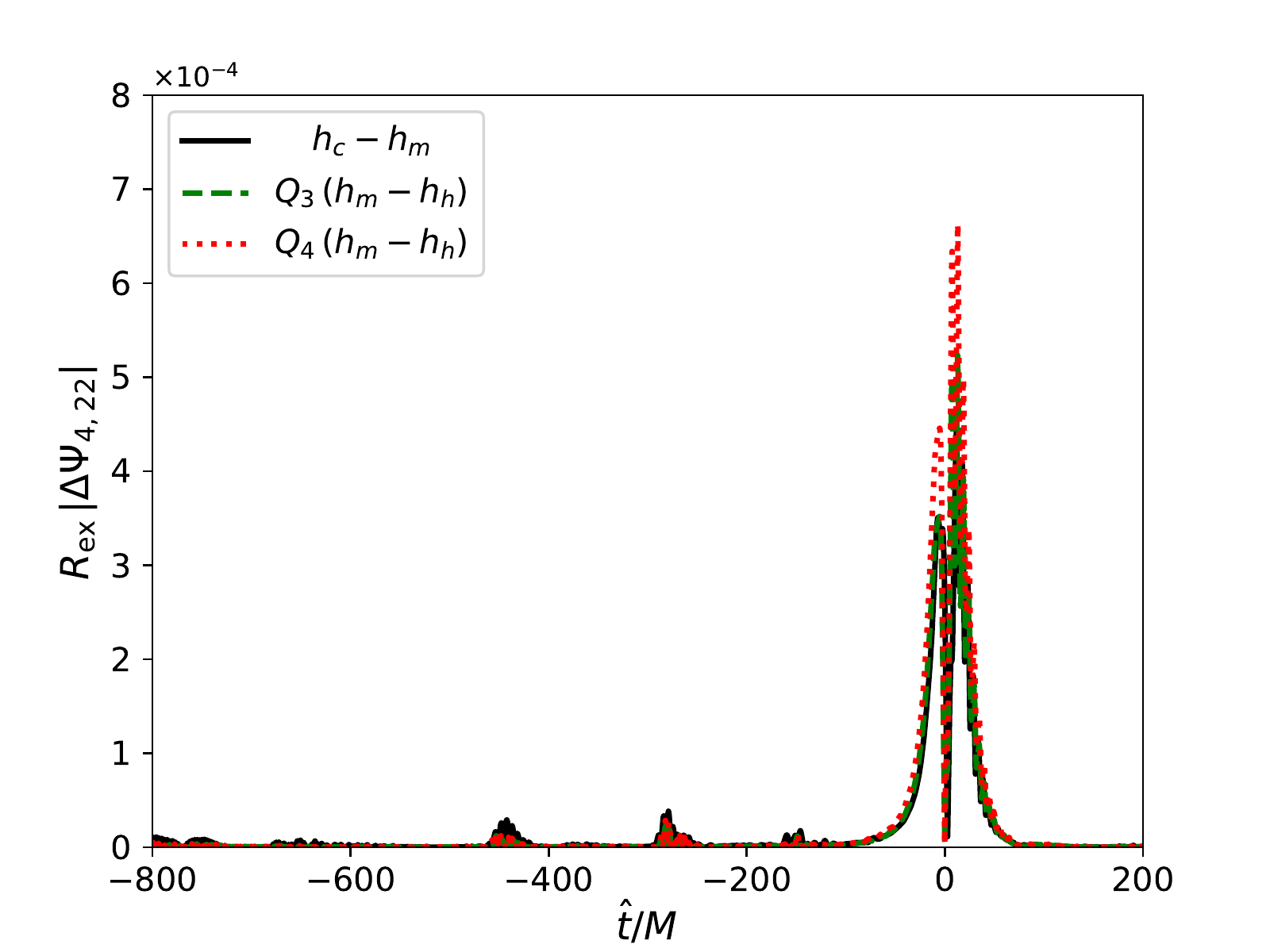}
\caption{\label{fig:Convergence_Psi4_l2m2_rex100}
(Color online) Convergence plot for run {\texttt{BBH\_q1}} showing the quadrupole of the gravitational waveform
  extracted at $\rex=100M$ and shifted in time such that $\hat{t}=0$ coincides with the time of merger.  We rescale the
  medium-high resolution waveform by $Q_{3}=0.56$ (green dashed line) and $Q_{4}=0.71$ (red dotted line) indicating
  third- to fourth-order convergence.  }
\end{figure}

\begin{table}[htpb!]
\caption{\label{tab:BBHQNMmodes} Postmerger ringdown frequencies of the scalar field obtained from a two-mode
  fit~\eqref{eq:KerrTwoModeFit}.  We list the scalar-led mode frequency $M_{f}\omega^{\rm S}_{lm}$ and the real part of
  the gravitational-led mode $M_{f} \omega^{\rm G}_{lm}$, rescaled by the final BH mass $M_{f}$.  We also denote the
  relative amplitude $\delta A_{\rm G} = A_{\rm G}/ A_{\rm S}$.
  A dash indicates that the ringdown is dominated by a single mode and we could not accurately estimate a secondary one.
} 
\begin{tabular}{|l|c|c|c|c|}
\hline
Run      & (lm)  & $M_{f} \omega^{\rm{S}}_{lm}$ & $M_{f} \omega^{\rm{G}}_{R,lm}$ & $\delta A_{\rm G}$ \\
\hline\hline
BBH\_q1  & (22)  & $0.64 - \imath 0.082$        & $0.53$                         & $3.3$          \\
         & (44)  & $1.15 - \imath 0.082$        & $1.03$                         & $0.2$          \\
\hline
BBH\_q12 & (11)  & $0.36 - \imath 0.094$        & $0.22$                         & $0.2$          \\
         & (22)  & $0.61 - \imath 0.084$        & $0.50$                         & $4.6$          \\
         & (33)  & $0.79 - \imath 0.094$        & --                             & --             \\
         & (44)  & --                           & $1.09$                         & --             \\
\hline
BBH\_q14 & (11)  & $0.33 - \imath 0.095$        & --                             & --             \\
         & (22)  & $0.57 - \imath 0.089$        & $0.45$                         & $0.5$          \\
         & (33)  & $0.74 - \imath 0.112$        & $0.65$                         & $0.7$          \\
         & (44)  & $0.98 - \imath 0.105$        & $0.92$                         & $0.5$          \\
\hline
\end{tabular}

\end{table}


\begin{figure}[htpb!]
\begin{center}
\includegraphics[width=0.5\textwidth,clip]{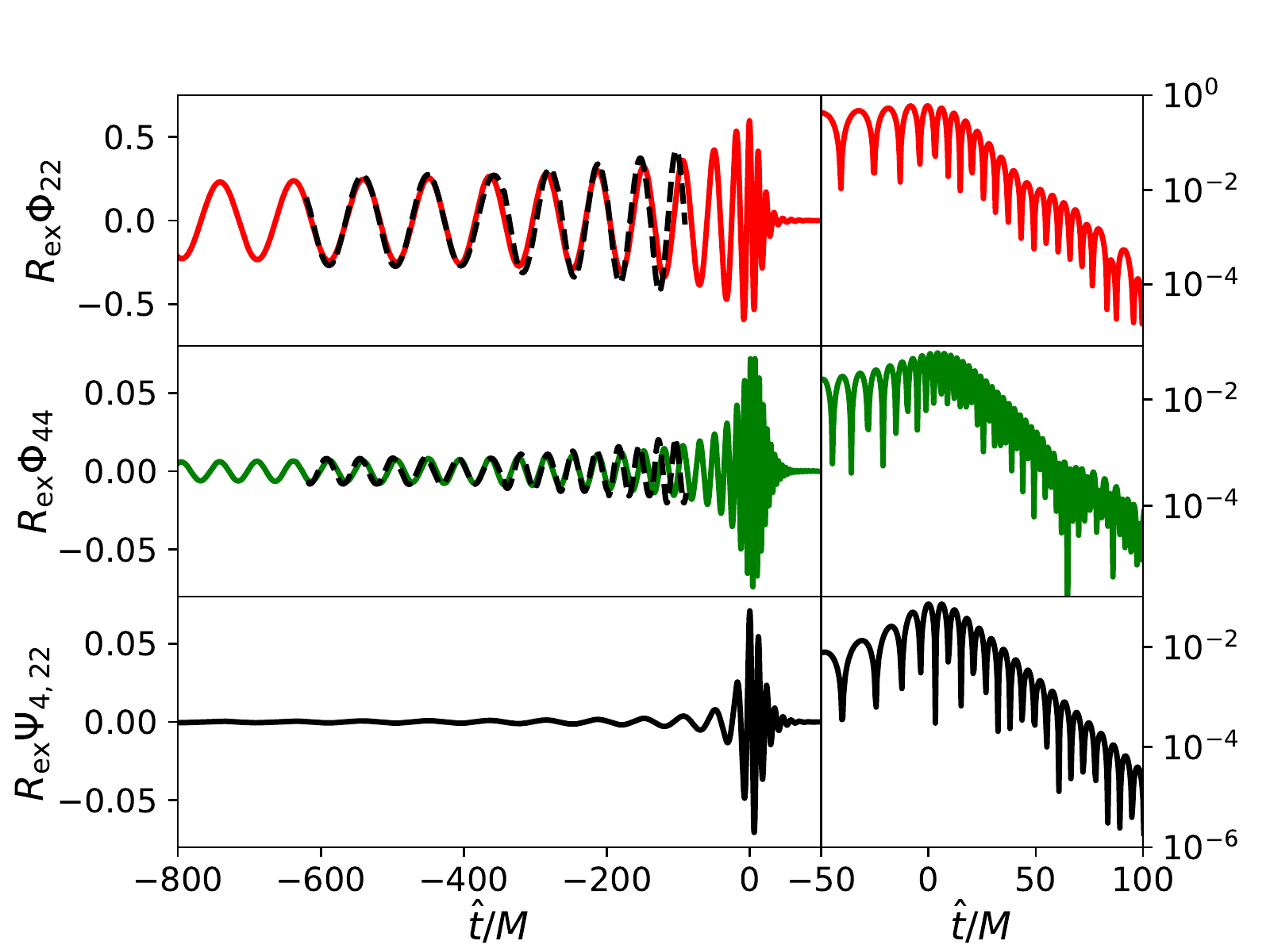}
\caption{\label{fig:BBH_q1_waveforms} (Color online) Scalar waveforms, rescaled by the extraction radius $\rex=100M$,
  sourced by an equal-mass, nonspinning BH binary whose waveform $\Psi_{4,22}$ is displayed in the bottom panel for
  comparison. $\hat{t}=0M$ indicates the merger time.  We show the $l=m=2$ (top panel) and $l=m=4$ (mid panel) modes of
  the scalar field.  During the inspiral phase we also display the PN waveform (black dashed lines, see
  Appendix~\ref{ssec:PNExpansion}).  In the right panels we zoom in on the merger-ringdown phase and observe a
  modulation due to the presence of both scalar-led and gravitational-led modes.  }
\end{center}
\end{figure}
\begin{figure*}[htpb!]
\begin{center}
\subfloat[$q=1/2$]{\includegraphics[width=0.5\textwidth,clip]{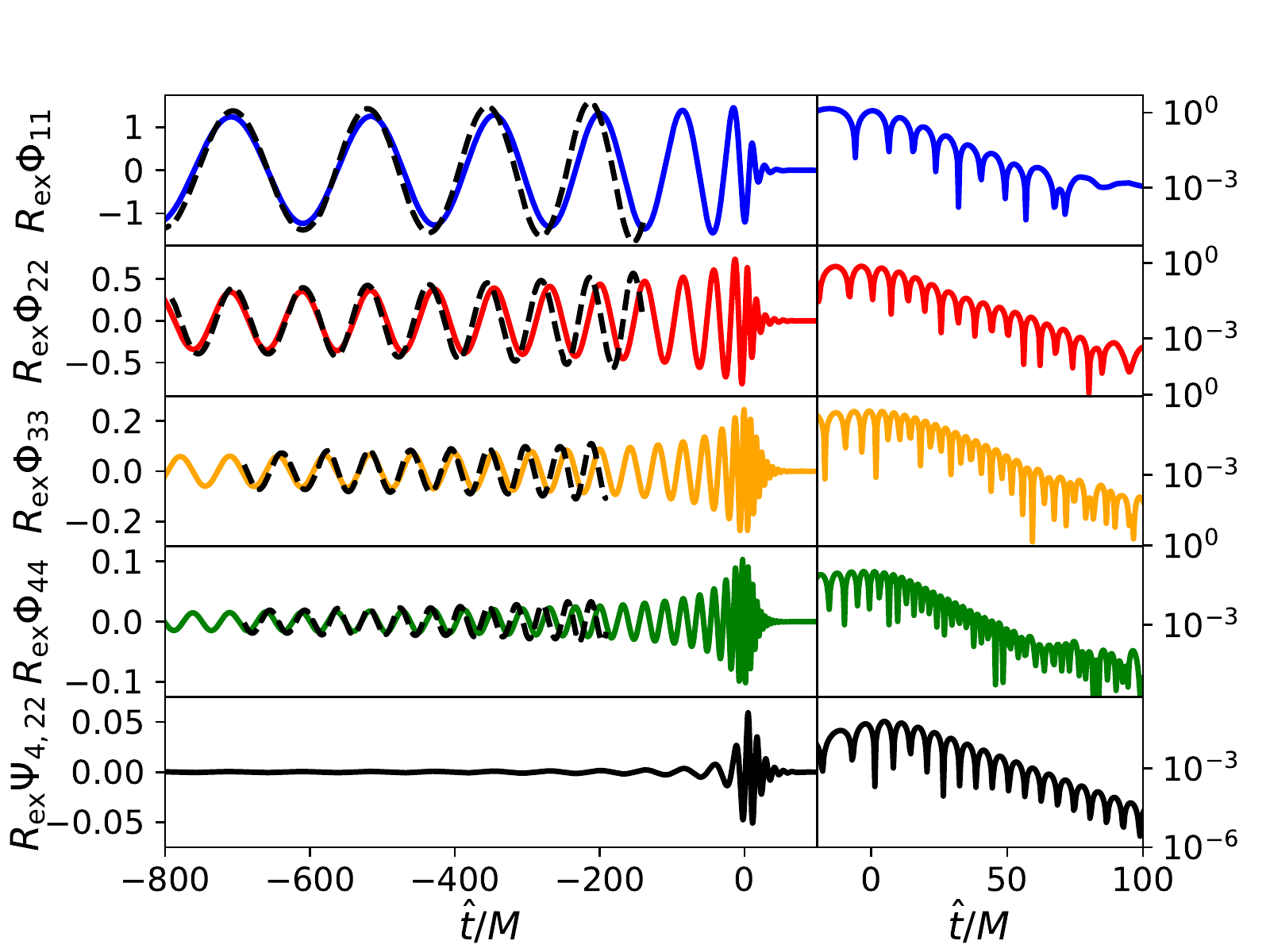}\label{fig:BBH_q12_waveforms}}
\subfloat[$q=1/4$]{\includegraphics[width=0.5\textwidth,clip]{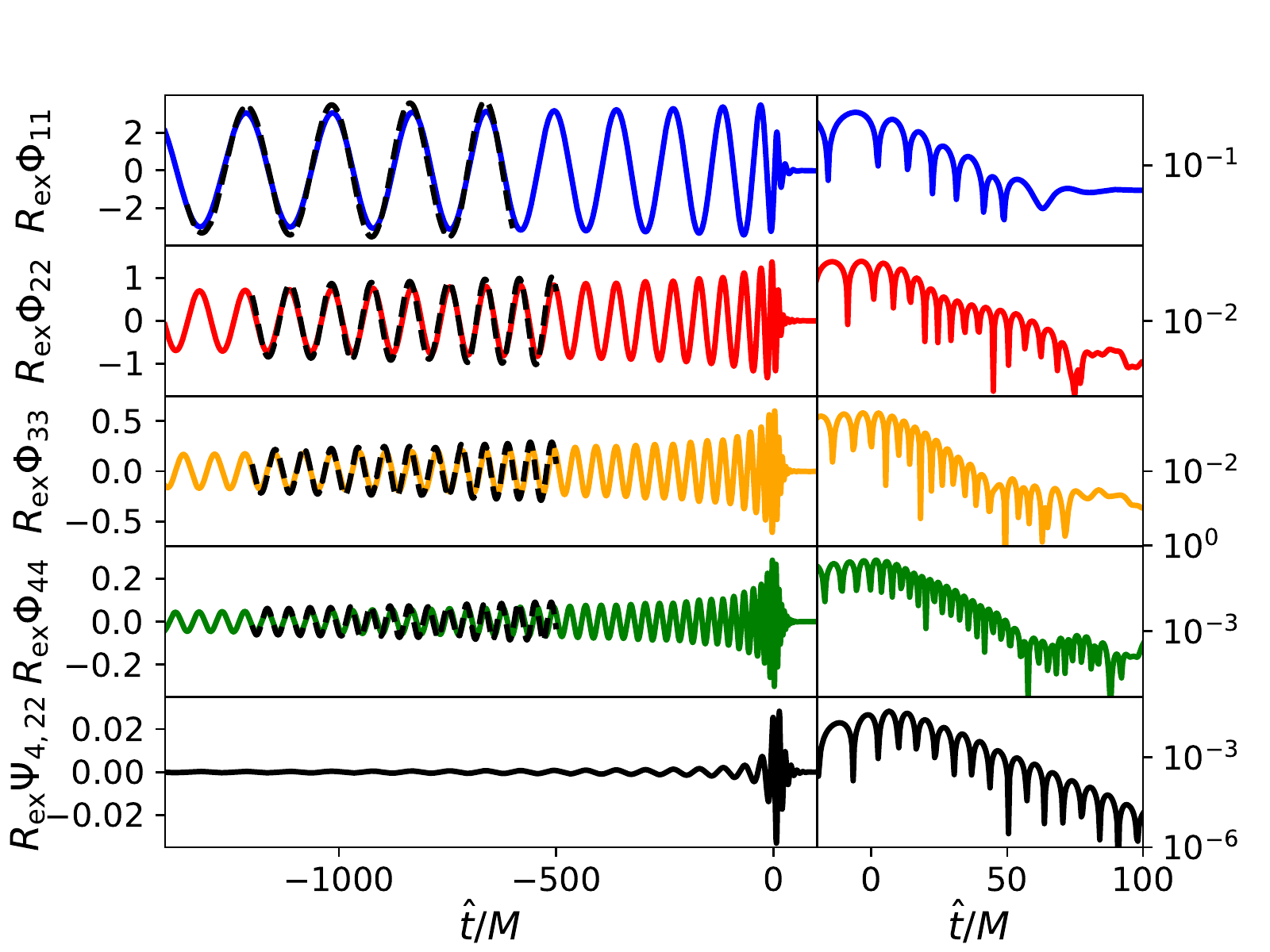}\label{fig:BBH_q14_waveforms}}
\caption{\label{fig:BBH_q12_q14_waveforms} (Color online) Same as Fig.~\ref{fig:BBH_q1_waveforms} but for $q=1/2$ (left
  panels) and $q=1/4$ (right panels).  In this case also the $l=m=1,3$ multipoles are emitted.  As before we compare the
  numerical data (solid lines) with the PN prediction (dashed lines).  }
\end{center}
\end{figure*}
\begin{figure}[htpb!]
\begin{center}
\includegraphics[width=0.5\textwidth,clip]{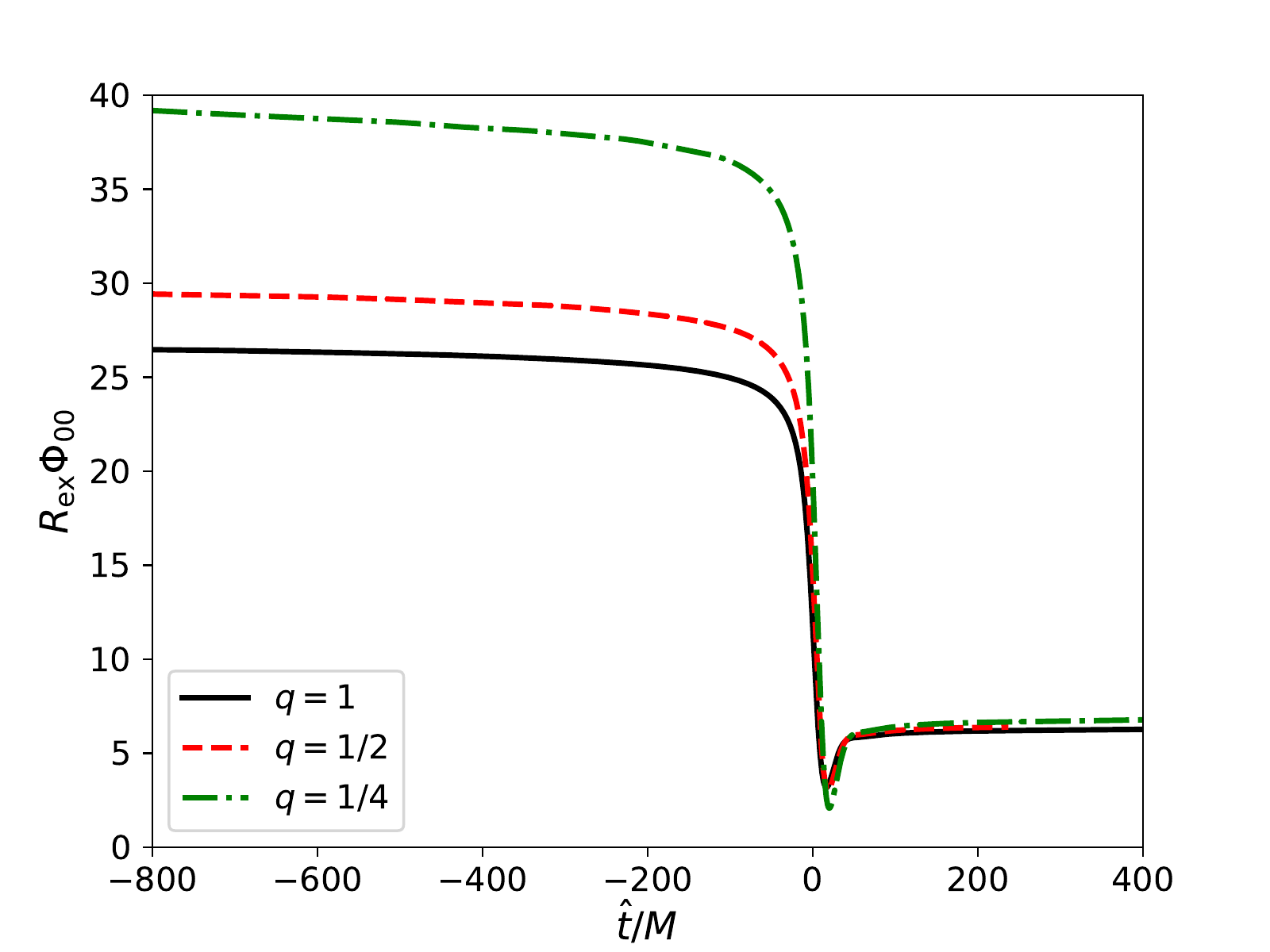}
\caption{\label{fig:BBH_Monopole_AllQ} (Color online) Same as Figs.~\ref{fig:BBH_q1_waveforms},
  \ref{fig:BBH_q12_q14_waveforms}, for the $l=m=0$ scalar mode.  The smallest mass ratio yields the largest pre-merger
  profile.  The final BHs have comparable masses and spin, and therefore similar scalar charge.  }
\end{center}
\end{figure}
To simulate these systems we set up a numerical grid with outer boundary at $256M$ consisting of $8$ refinement levels.
For the results presented here we used a grid spacing of dx$=1.6M$ in the outermost refinement level.  To validate our
results we performed a convergence analysis of run {\texttt{BBH\_q1}} using additional grid spacings of dx$_{c}=1.8M$
and dx$_{h}=1.0M$.  We estimate the numerical error to be about
\begin{enumerate*}[label={(\roman*)}]
\item $\Delta\Psi_{4,22}/\Psi_{4,22} \lesssim1.5\%$ in the gravitational waveforms;
\item $\Delta\Phi_{22}/\Phi_{22} \lesssim1.5\%$ in the scalar waveforms;
and
\item $\lesssim0.5\%$ in both the gravitational and scalar phases.
\end{enumerate*}
The corresponding convergence plot is shown in Fig.~\ref{fig:Convergence_Psi4_l2m2_rex100}.

{\textbf{Waveforms:}}
We present the background gravitational waveform 
and the ${\cal O}(\epsilon)$ scalar waveform for binaries with mass
ratio $q=1$ and $q=1/2,1/4$ in Figs.~\ref{fig:BBH_q1_waveforms} and \ref{fig:BBH_q12_q14_waveforms}, respectively.
All presented waveforms are shifted in time such that 
$\hat{t}/M=(t-\tM-\rex)/M=0$ indicates the maximum in the dominant 
gravitational mode as measure for the time of merger.
The waveforms exhibit the typical morphology: a sinusoid with increasing frequency that is driven by the orbital motion
of the BHs, the highly nonlinear merger followed by the exponentially damped ringdown.
During the inspiral we compare the numerical results to the analytical expressions 
obtained at leading PN
%
order in~\cite{Yagi:2011xp} (see Appendix~\ref{ssec:PNExpansion}). We remark that these expressions depend on the
time-dependent orbital frequency $\Omega(t)$, which, at this PN order, can not be obtained with good
approximation. Therefore, we extract the orbital frequency from the numerical data (measuring, at each half-cycle, the
wavelength of the gravitational waveform). Within this approach, which is similar to that used
in~\cite{Okounkova:2017yby}, the comparison between PN and numerical results concerns the {\it amplitudes} of the scalar
waveforms, while their phases agree by definition.

Interestingly, while the scalar signal for $l\geq2$ is qualitatively similar to the gravitational waveform and displays
the classical chirp, the dipole is qualitatively different. As shown in Fig.~\ref{fig:BBH_q12_q14_waveforms}, the
frequency of the dipole mode grows as expected during the merger, but the amplitude remains almost constant.
This is a strong-field behavior that is not captured by the PN approximation.  A potential explanation of this behaviour
is that the scalar configuration ceases to be dominantly dipolar before the merger, {\em i.e.}~the dynamical evolution
of the scalar is more complex and it involves additional oscillations and reconfiguration. Our simulations, and in
particular the time evolution of the scalar distribution, do seem to be consistent with this explanation, though
limitations in resolution do not allow us to make a conclusive statement.

In the post-merger phase the background approaches a stationary spinning BH, so we expect to observe the same multiple
ringing discussed in the previous section for an isolated BH. This is confirmed in the insets of
Figs.~\ref{fig:BBH_q1_waveforms} and \ref{fig:BBH_q12_q14_waveforms} and by the postmerger ringdown frequencies
extracted from the scalar waveform using the two-mode fit~\eqref{eq:KerrTwoModeFit} and presented in
Table~\ref{tab:BBHQNMmodes}.  Note that, in contrast to the single BH case, the background is now a perturbed BH plus
gravitational radiation, both of which modify the source term of the scalar field. In particular, gravitational
radiation seems to cause an enhancement of the gravitational-led quadrupole modes, which dominate over the scalar-led
one in some configurations (see, e.g., the $l=m=2$ case for $q=1,1/2$ in Table~\ref{tab:BBHQNMmodes}).

Finally, the scalar field monopole for the same values of the mass ratio is shown in
Fig.~\ref{fig:BBH_Monopole_AllQ}. The pre-merger amplitude is larger for smaller values of the mass ratio $q$, while the
final amplitude is approximately independent of $q$. This behaviour can be understood noting that the post-merger
amplitude is, as a first approximation, $\Phi\simeq\alpha_{\rm{GB}}/(2Mr)$ (see App.~\ref{sssec:SmallSpinApprox}), where
(by construction) the total mass $M$ is the same in all cases, {\em i.e.}, independent of the mass ratio.  Instead, when
the two BHs are well separated, the scalar field amplitude is
\begin{align}
\label{eq:PhiPreMerger}
\Phi\simeq &  \frac{\alpha_{\rm{GB}} }{2 m_{1} r } + \frac{\alpha_{\rm{GB}} }{2 m_{2} r }
        =  \frac{\alpha_{\rm{GB}} }{2 M r } \frac{1}{\eta}
\,,
\end{align}
for sufficiently large radii encompassing the entire binary.  Therefore, the ratio between the pre-merger and the
post-merger amplitude is expected to be determined by the (inverse of the) symmetric mass ratio $\eta$.  In particular
we have $1/\eta=4$,~$4.5$ and~$6.25$ for mass ratios $q=1$,~$1/2$ and~$1/4$.  These values are in agreement with
Fig.~\ref{fig:BBH_Monopole_AllQ}.

{\textbf{Energy and momentum fluxes:}} Next, we investigate the energy radiated in gravitational and scalar waves.  We
compute their energy fluxes using~\eqref{eq:FluxesGWGR}--~\eqref{eq:FluxesGeneral} with~\eqref{eq:DefFluxes2ndOrderSum},
{\em i.e.} accounting for both the canonical scalar's and Gauss--Bonnet contributions to the energy flux.
We furthermore estimate the total radiated energy by integrating Eqs.~\eqref{eq:FluxesGWGR} and~\eqref{eq:FluxesGeneral}
in time, measuring it at different extraction radii and performing the extrapolation
\begin{align}
E^{\rm{GW}}/M = & E^{\rm{GW}}_{\infty}/M + B / \rex 
\,,
\end{align}
and likewise for the second-order flux transported by the scalar field.  We estimate the extrapolation error by
comparing to $E/M = E_{\infty} + B / \rex + C / \rex^{2}$ and find $\Delta E^{\rm{GW}}_{\infty} / E^{\rm{GW}}_{\infty}
\lesssim 0.8\%$ and $\Delta E^{\rm{(2)}}_{\infty} / E^{\rm{(2)}}_{\infty} \lesssim 7\%$.
The results are summarized in Table~\ref{tab:BBHSetup}.

\begin{figure*}[htpb!]
\includegraphics[width=0.32\textwidth,clip]{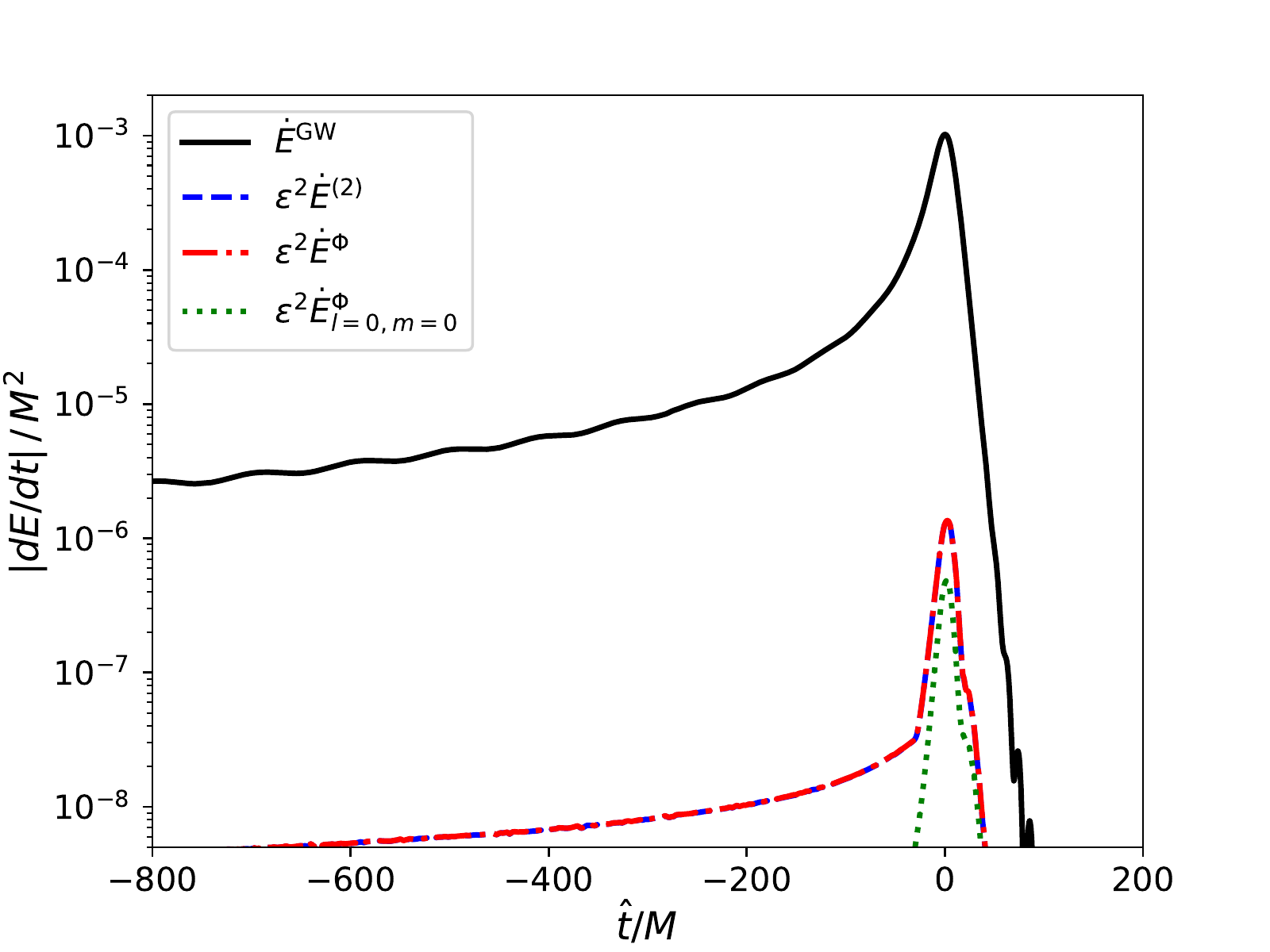}
\includegraphics[width=0.32\textwidth,clip]{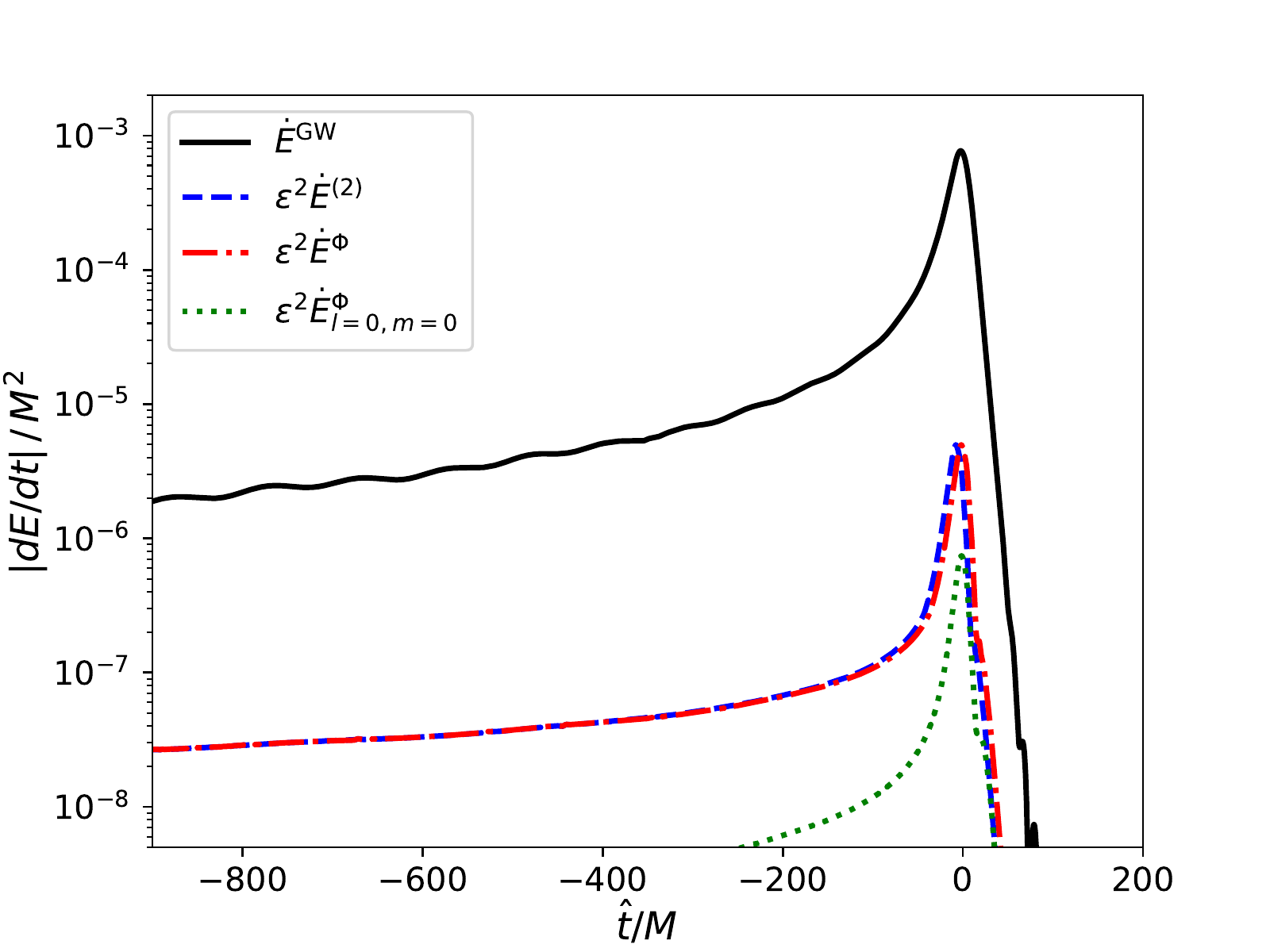}
\includegraphics[width=0.32\textwidth,clip]{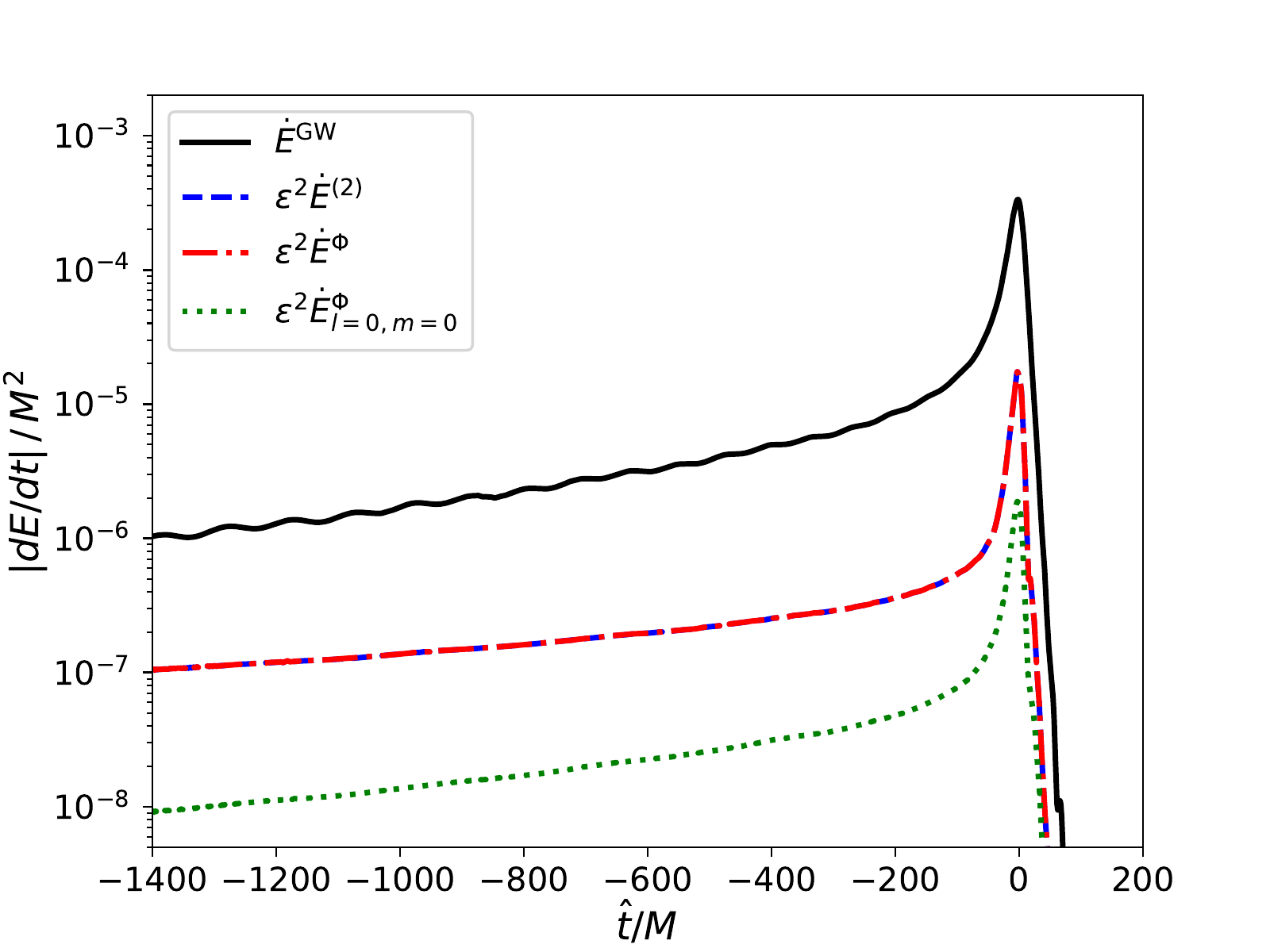}
 \caption{\label{fig:BBH_Eflux_Psi4_Phi} (Color online) Gravitational (black solid lines) and scalar second-order (blue dashed
   lines) energy fluxes measured at $\rex=100M$ for mass ratios $q=1$ (left), $q=1/2$ (middle) and $q=1/4$ (right). They
   have been shifted in time such that $\hat{t}=0$ coincides with the time of merger.  We rescaled the scalar field flux
   by the appropriate powers of the expansion parameter, and set $\epsilon=0.01$.  We also show the canonical
   scalar field flux (red dot-dashed lines) and its monopole contribution (green dotted lines) that
   exhibits a peak during the merger when the scalar field adjusts to the final, single BH solution.  }
\end{figure*}

In Fig.~\ref{fig:BBH_Eflux_Psi4_Phi} we present the fluxes for all three configurations.
The background, {\em i.e.}, GW flux (black solid lines) follows the common pattern: it increases monotonically in
amplitude as the BHs circle around each other for the last few orbits, culminates in a peak during their merger, and
decays exponentially as the newly born BH rings down to a Kerr BH.
We also show the second-order energy flux carried by the scalar waves~\eqref{eq:DefFluxes2ndOrderSum} (blue dashed lines) 
together with the canonical scalar field energy flux ${\dot E}^{(\Phi)}$ (red dot-dashed lines), rescaled by the appropriate power
$\epsilon^{2}$ of the expansion parameter.  Exemplarily, we set $\epsilon=0.01$.  We observe that the second-order scalar flux
is dominated by the contribution of the scalar stress-energy tensor.

The morphology of the signal is determined by the orbital dynamics and monotonically increases during the inspiral of
the background BHs. The canonical scalar field flux also exhibits a peak during the merger that is predominantly
determined by the monopole mode, as is illustrated by the green dot-dashed lines in Fig.~\ref{fig:BBH_Eflux_Psi4_Phi}.
This is because the system changes rapidly from two nonrotating BHs, each with its own scalar hair determined
by~\eqref{eq:PhiPreMerger} to a single rotating BH with a new scalar configuration of this form but with larger mass
and, hence, smaller scalar charge.

The ratio between scalar and gravitational radiation dramatically increases as the mass ratio decreases.  This is
because the scalar charge is determined by the smallest mass scale in the system yielding the largest curvatures, and
then undergoes a transition to the final BH mass.  So while this characteristic scale changes at most by a factor of two
in the equal-mass case, it can be vastly different as we decrease the mass ratio.
Whether and how this trend continues for higher mass ratios is beyond the scope of the paper and will be presented in a
more detailed parameter study elsewhere.

\section{Range of validity and observational constraints}\label{sec:range}
Our results allow us to put new constraints on the Gauss--Bonnet coupling with previous and upcoming GW detections.
However, before doing so we need to quantify the validity of the low-energy perturbative expansion presented in
Sec.~\ref{ssec:EFTapproach} up to first order.
Therefore, we consider the ``instantaneous'' range of validity as well as integrated, secular effects. In the former
case we demand that the scalar energy flux at a given instant $t$ be much smaller than the GW flux, whereas in the
latter case we require that the dephasing due to scalar emission accumulated during the inspiral be smaller than the GW
phase.
Although we have not explicitly evolved the second-order scheme we can estimate deviations at this order from the source
terms in Eqs.~\eqref{eq:ExpansionEps2Phi0Metric} as they only depend on background or first-order quantities that we
obtained in our numerical simulations.

\subsection{Instantaneous range of validity}~\label{ssec:InstantValidity}
We start by investigating the instantaneous range of validity, that is, we check for which couplings the perturbative
expansion~\eqref{eq:expansionSFMetric} remains applicable at every timestep in our simulation.
To this end we compare the second-order energy flux (carried by the scalar waves) with the background GW flux. Then, a
necessary condition for the perturbative expansion to apply is
\begin{align}
\dot E^{\rm GW}\gg & \frac{1}{2}\epsilon^2 \dot E^{(2)}
\,.
\end{align}

\begin{figure}[b]
\begin{center}
\includegraphics[width=0.5\textwidth,clip]{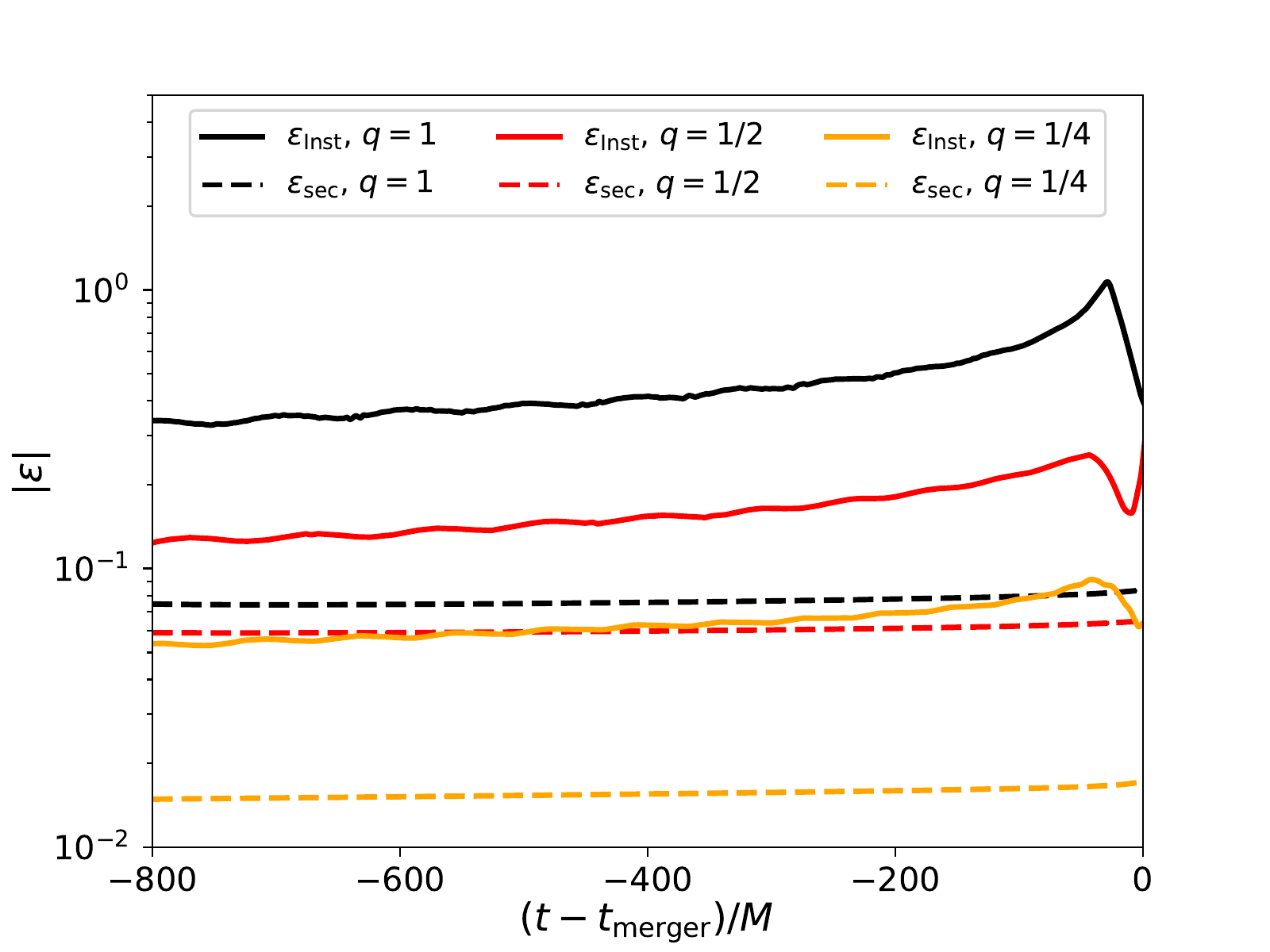}
\caption{\label{fig:BBH_Compare_EpsEflux_Dephasing_allq} (Color online) Comparison between the instantaneous (solid
  lines) and cumulative (dashed lines) ranges of validity for all considered mass ratios.  }
\end{center}
\end{figure}

In Fig.~\ref{fig:BBH_Compare_EpsEflux_Dephasing_allq} we present the (instantaneous) bounds on the dimensionless
coupling obtained from
\begin{align}
\label{eq:convergence2}
\normfct{\epsilon} \lesssim & 
\sqrt{2\frac{\dot E^{\rm GW}}{\dot E^{(2)}}}
\,.
\end{align}
We observe that the tightest constraints come from binaries with small mass ratio, because in such case the first of the
binary components has a smaller mass than in the equal-mass case (recall that the total mass is fixed to unity) and
therefore yields a larger dimensionless coupling $\sim\alpha_{\rm GB}/m_1^2$.  This is consistent with the fact that sGB
gravity is a strong-curvature correction to GR, so the strongest effects come from small BHs for which the near-horizon
curvature is large.

At the merger nonlinear effects dominate, the scalar field transitions to a new configuration (see
Fig.~\ref{fig:BBH_Monopole_AllQ}) while the final BH forms, and we observe a burst of scalar energy flux.  This is
indicated by the dip around $\hat{t}=0$ in Fig.~\ref{fig:BBH_Compare_EpsEflux_Dephasing_allq}, where the allowed value
of the dimensionless coupling drops by about an order of magnitude.

\subsection{Secular effects and dynamical range of validity}~\label{ssec:SecEffectsValidity}
While the instantaneous range of validity is a first check, we find it more instructive to explore the influence of the
scalar field on the binary's evolution.
As we have seen in Sec.~\ref{sec:NumericalResults} the (background) spacetime dynamics source scalar radiation that
needs to be accounted for in the full energy budget.  Because a BH binary in sGB gravity emits not only gravitational
but also scalar waves, the inspiral is accelerated as compared to the same system in GR.  This yields an increase in the
orbital angular velocity of the binary and causes a dephasing $\Delta\phi$ in the gravitational waveform when compared
to GR.
Since we adopted a perturbative approach up to first order only,
we cannot compute this dephasing -- a second-order effect -- directly, but we can provide qualitative estimates based on our
numerical results.

In accordance with this approach we expand the orbital phase $\phi$ and orbital frequency $\Omega=\dot\phi$ as in
Eq.~\eqref{eq:expansionGeneral}.  Recalling that the first-order contribution to the metric vanishes [see
  Eq.~\eqref{eq:SolFirstOrder}] we note that there is also no first-order correction to the orbital phase and frequency,
and the first nonvanishing shifts appear at second order in the perturbative expansion, {\em i.e.},
\begin{subequations}
\begin{align}
\label{eq:ExpansionPhaseFreq}
\phi \simeq \,& \phi^{(0)} + \frac{1}{2}\epsilon^{2} \phi^{(2)} + 
\Ord{\epsilon}{3}
\,,\\
\Omega \simeq \,&  \Omega^{(0)} + \frac{1}{2}\epsilon^{2} \Omega^{(2)} + 
\Ord{\epsilon}{3}
\,,
\end{align}
\end{subequations}
where $\phi^{(0)}$ and $\Omega^{(0)}=\lvert\vec{d}\times\dot{\vec{d}} \rvert / d^{2}$ are the orbital phase and
frequency of the BH binary calculated from the distance $\vec{d}$ between the puncture positions, and
$\Delta\phi=\epsilon^2\phi^{(2)}/2$, $\Delta\Omega=\epsilon^2\Omega^{(2)}/2$ are the Gauss--Bonnet corrections to the
phase and frequency.
The validity of the perturbative expansion requires $\phi^{(0)}\gg \Delta\phi$, where
\begin{align}
\label{eq:PhiIntOmega}
 \phi^{(i)}(t) = & \int_0^t \dif t' \Omega^{(i)}(t')\,.
\end{align}
Therefore, the threshold is
\begin{align}
\label{eq:condepssecular}
\lvert\epsilon(t)\rvert \lesssim &\sqrt{2\frac{\phi^{(0)}(t)}{\phi^{(2)}(t)}}\,. 
\end{align}
We evaluate $\phi^{(i)}(t)$ through
\begin{equation}
\label{eq:ddphi2}
\dot  \phi^{(0)} =  \Omega^{(0)}
\,,\quad
\ddot \phi^{(2)} =   \dot\Omega^{(2)}
\,,
\end{equation}
\begin{align}
\label{eq:DOmega2DtGen}
\dot\Omega^{(2)} = & \frac{\dot\Omega^{(0)}}{\dot E^{(0)}}
        \left[ \dot E^{(2)} - \dot\Omega^{(0)} \frac{dE^{(2)}}{d\Omega}
        \right] 
\,,
\end{align}
where $\dot E^{(0)}=\dot E^{\rm{GW}}$ and $\dot E^{(2)}$ are the GW flux in GR and the second-order energy flux carried
by the scalar waves, respectively.  While these are computed from the numerical data using Eqs.~\eqref{eq:EnergyFluxGWGR}
and~\eqref{eq:DefFluxes2ndOrderSum}, we can only estimate the last term ${dE^{(2)}}/{d\Omega}$ in
Eq.~\eqref{eq:DOmega2DtGen} using the PN approximation.

As shown in~\cite{Stein:2013wza}, the scalar interaction binding energy of a compact binary is
\begin{equation}
  E_{bind}=\frac{(-1)^t}{4}(2s+2t-1)\mu_{1}^{i_1\dots i_s}\mu_{2}^{j_1\dots 
j_t}\frac{n_{12}^{i_1\dots i_sj_1\dots j_t}}{r^{s+t+1}}
  \label{eq:binding-steinyagi}
\end{equation}
where $s,t$ are the (leading) multipole numbers of the scalar field emission from the two BHs, and $\mu_1$, $\mu_2$ are
their multipole charges (see App.~\ref{app:CouplingConventions} for the different notations for charges). Note that,
while in the case of Chern-Simons gravity studied in~\cite{Okounkova:2017yby} the leading-order contribution to the
binding energy is the dipole-dipole interaction (only present when the BHs are rotating), in the case of sGB gravity the
leading-order binding energy contribution is the monopole-monopole interaction $E_{MM}$, which does not depend on the BH
spins. Therefore, $s=t=0$ in Eq.~\eqref{eq:binding-steinyagi}, and 
\begin{equation}
  \frac{1}{2}\epsilon^2E^{(2)}\simeq E_{MM}=-\frac{1}{4}
  \frac{\mu_1\mu_2}{r}=-\frac{1}{4}
\frac{\mu_1\mu_2\Omega^{2/3}}{M^{1/3}}\label{eq:e2}
\end{equation}
where $\mu_i=\alpha_{GB}/(2m_i)=2\epsilon M^2/m_i$ is the scalar charge (see~\cite{Yagi:2011xp} and
App.~\ref{sssec:SmallSpinApprox}) of the $i$-th body, and, at leading PN order, $\Omega=(M/r^3)^{1/2}$~\footnote{Note
  that Eq.~\eqref{eq:e2} implies that the monopole-monopole scalar interaction is attractive (as long as $\mu_1\mu_2>0$,
  as in the case of sGB BHs).}. Therefore,
\begin{equation}
E^{(2)}=-\Omega^{2/3}M^{5/3}\frac{M^2}{m_1m_2}
\end{equation}
and
\begin{equation}
  \frac{dE^{(2)}}{d\Omega}\simeq-\frac{4}{3} (\Omega^{(0)} )^{-1/3}M^{5/3}\frac{(1+q)^2}{q}\,.\label{eq:dE2dO}
\end{equation}
Now we have all the ingredients to compute the condition~\eqref{eq:condepssecular} for the expansion to be consistent
also during the binary's evolution.  We illustrate it in Fig.~\ref{fig:BBH_Compare_EpsEflux_Dephasing_allq} as function
of time (shifted by the time of merger).

As in the previous case, the bounds on the validity of the perturbative approach up to first order become more stringent
with decreasing mass ratio and near the merger, as one might expect in this highly nonlinear regime.

\subsection{Observational bounds}
While an accurate computation of the observational constraints on the Gauss--Bonnet coupling would require computing the
${\cal O}(\epsilon^2)$ corrections to the gravitational waveform (which is a higher-order effect than the ones computed
in this work), we can estimate an observational constraint based on the assumption that no GW dephasing
$\DPGW = 2\Delta\phi=\epsilon^2\phi^{(2)}+{\cal O}(\epsilon^3)$ induced by some non-GR extension is observed by a given (present
or future) GW detector and, hence, it must be at least below the detector's GW phase 
uncertainty $\DPdet$.  
This is a conservative estimate since: (i)~smaller effects could be constrained by comparing 
directly waveform models in GR and in modified
theory; and (ii)~the contribution of the energy flux at ${\cal O}(\epsilon^2)$ due to 
gravitational radiation, which we are neglecting,
would further increase the dephasing. 
Computing the full energy flux to ${\cal O}(\epsilon^2)$ is an interesting problem which 
we leave for future work. 

We remark, however, that in this estimate we are not taking into account $O(\epsilon^2)$ shift in the physical masses of
BHs in sGB gravity. For instance, in EdGB gravity the physical mass of an isolated, 
static BH measured from the asymptotic limit of the metric is 
${\cal  M}\sim M(1+49\alpha_{\rm GB}^2/(20480\pi M^4))$~\cite{Mignemi:1992nt,Kanti:1995vq,Yunes:2011we}.  
Since the orbital energy $-Gm_1m_2/(2r)$ and the monopole
scalar binding energy~\eqref{eq:e2} have similar expressions, the mass shift and the scalar monopole energy may have
degenerate effects~\footnote{We thank Leo Stein for pointing this out to us.}.
On the other hand, we have verified that neglecting the contribution of the monopole binding energy
$\frac{dE^{(2)}}{d\Omega}$ in Eq.~\eqref{eq:DOmega2DtGen} would modify the dephasing 
by $\lesssim20\%$.
Moreover, the dissipative and conservative contributions to the dephasing (i.e. the two terms in  Eq.~\eqref{eq:DOmega2DtGen})
have different dependence on the parameters of the binary system.
Thus, unless
fine-tuning cancelations occur, our computations provide a correct order-of-magnitude estimate of the dephasing due to
sGB gravity. 

Absence of non-GR dephasing appears to be the case in previous LIGO
detections~\cite{TheLIGOScientific:2016src,Abbott:2017oio}, which can therefore be used to put actual upper
bounds. Using the same assumption of a negative search, we can also forecast observational bounds for third-generation
detectors such as the Einstein Telescope or Cosmic Explorer and for the space-based LISA mission.
This can be translated into the bound
\begin{align}
\label{eq:BoundEpsNonDet}
\epsilon \lesssim & \sqrt{\frac{\DPdet}{\phi^{(2)}} }
\,,
\end{align}
where we again evaluate $\phi^{(2)}$ as discussed in the previous section. 
Specifically, we consider the reference systematic phase errors
\begin{enumerate*}[label={(\roman*)}]
\item $\Delta\phi_{\rm{LIGO}}\lesssim0.1 $ for LIGO's O2 
run~\cite{Abbott:2017oio};
\item $\Delta\phi_{\rm{3G}} \lesssim0.01$ for third generation ground-based detectors~\cite{Essick:2017wyl}; and
\item $\Delta\phi_{\rm{LISA}}\lesssim0.01$ for 
LISA~\cite{Berti:2004bd,Audley:2017drz}. 
\end{enumerate*}

While LIGO allows us to constrain the dimensionless Gauss--Bonnet coupling to be $\lesssim\Ord{10}{-3}$,
third-generation ground-based detectors and the space-based LISA mission will provide bounds that are about one order of
magnitude more stringent. Note, however, that since $\alpha_{GB}\sim\epsilon M^2$ the bounds become much weaker for
heavier BHs, so supermassive BHs are not good probes of higher-curvature corrections to GR.

\begin{figure}[b] 
\begin{center}
\includegraphics[width=0.5\textwidth,clip]{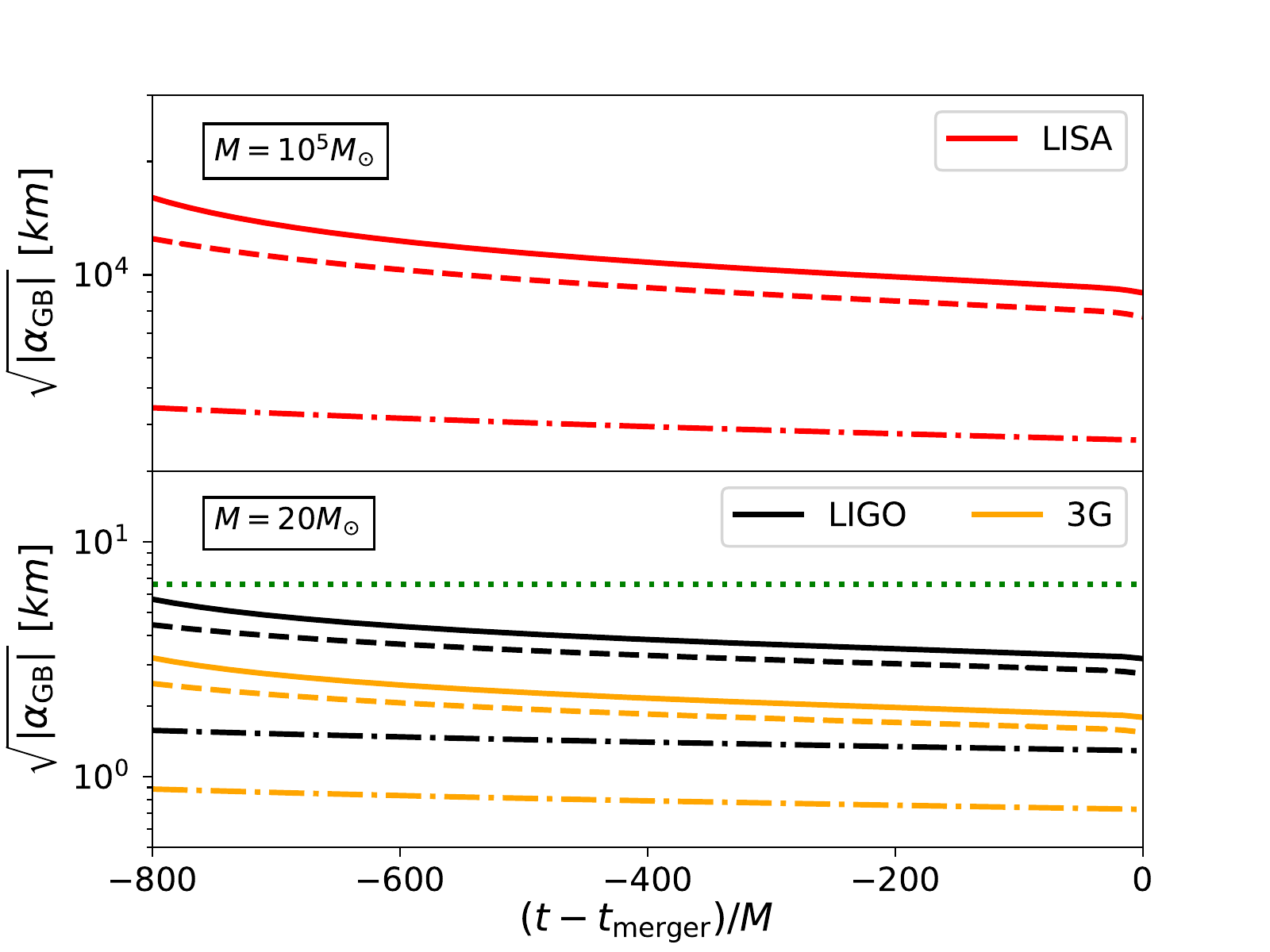}
\caption{\label{fig:BBH_AlpDet_vs_time_allq} 
(Color online) Bounds on dimensionful coupling parameter $\alpha_{\rm{GB}} = 4 \epsilon M^{2}$ inferred from the
  detector bound~\eqref{eq:BoundEpsNonDet} and, exemplarily, considering a GW151226-type source.
We present estimates for LIGO's O2 run (black lines) and forecasts for third generation ground-based detectors (orange
lines) and LISA (red lines). We consider all simulated mass ratios $q=1$ (solid lines), $q=1/2$ (dashed lines) and
$q=1/4$ (dashed-dotted lines).
For comparison we also show the so far most stringent bound coming from the existence of light BHs (green dotted line).
}
\end{center}
\end{figure}

This can be seen in Fig.~\ref{fig:BBH_AlpDet_vs_time_allq}, where we present bounds on the {\textit{dimensionful}} GB
coupling that depends on the mass of the system.
We choose a binary with $M=20M_{\odot}$ for ground-based detectors and $M=10^{5}M_{\odot}$ for LISA. 
Furthermore, we
summarize the constraints on both the dimensionless and dimensionful coupling constants in
Table~\ref{tab:ConstraintsForecast}.
For ground-based detectors, we consider total binary masses of $M=20M_{\odot}$ or $M=60M_{\odot}$, corresponding
to the lightest and most massive BH binary detected so far, whereas we consider $M=10^{5}M_{\odot}$ for space-based
detectors.

Using this phase information we can put the most stringent 
observational constraints on the Gauss--Bonnet coupling to
date.  We choose to consider a binary with mass ratio $q=1/2$ and
$M\sim20M_{\odot}$ because its characteristics strongly resemble X
GW151226~\cite{Abbott:2016nmj}. Hence, our results, within the caveats of our 
approach, place the constraint~\footnote{ We recall (see Sec.~\ref{sec:intro}) 
that the low mass -ray binary
  constraint $\sqrt{\lvert\alpha_{\rm{YYSP}}\rvert}\lesssim 1.9$km~\cite{Yagi:2012gp} and GW constraint
  $\sqrt{\lvert\alpha_{\rm{YYSP}}\rvert}\lesssim 5.1$km~\cite{Yunes:2016jcc} from GW151226 translate to
  $\sqrt{\lvert\alpha_{\rm{GB}}\rvert}\lesssim 10$km and $\sqrt{\lvert\alpha_{\rm{GB}}\rvert}\lesssim 27$km,
  respectively, in our notation; see Eq.~\eqref{eq:TrafoCouplingYYSPvsKanti}.}
\begin{align}
\label{eq:boundLIGOGW151226}
\sqrt{\alpha_{\rm{GB}}}\lesssim & 2.7 \,{\rm km}
\,.
\end{align}
This bound can improve with detections of less massive systems or binaries with a lower mass ratio. 
As indicated in Table~\ref{tab:ConstraintsForecast} and Fig.~\ref{fig:BBH_AlpDet_vs_time_allq}, upcoming
third-generation ground-based detectors will have the potential to place even more stringent constraints on
Gauss--Bonnet-type modifications to GR.

\begin{figure}[htpb!]
\begin{center}
\includegraphics[width=0.5\textwidth,clip]{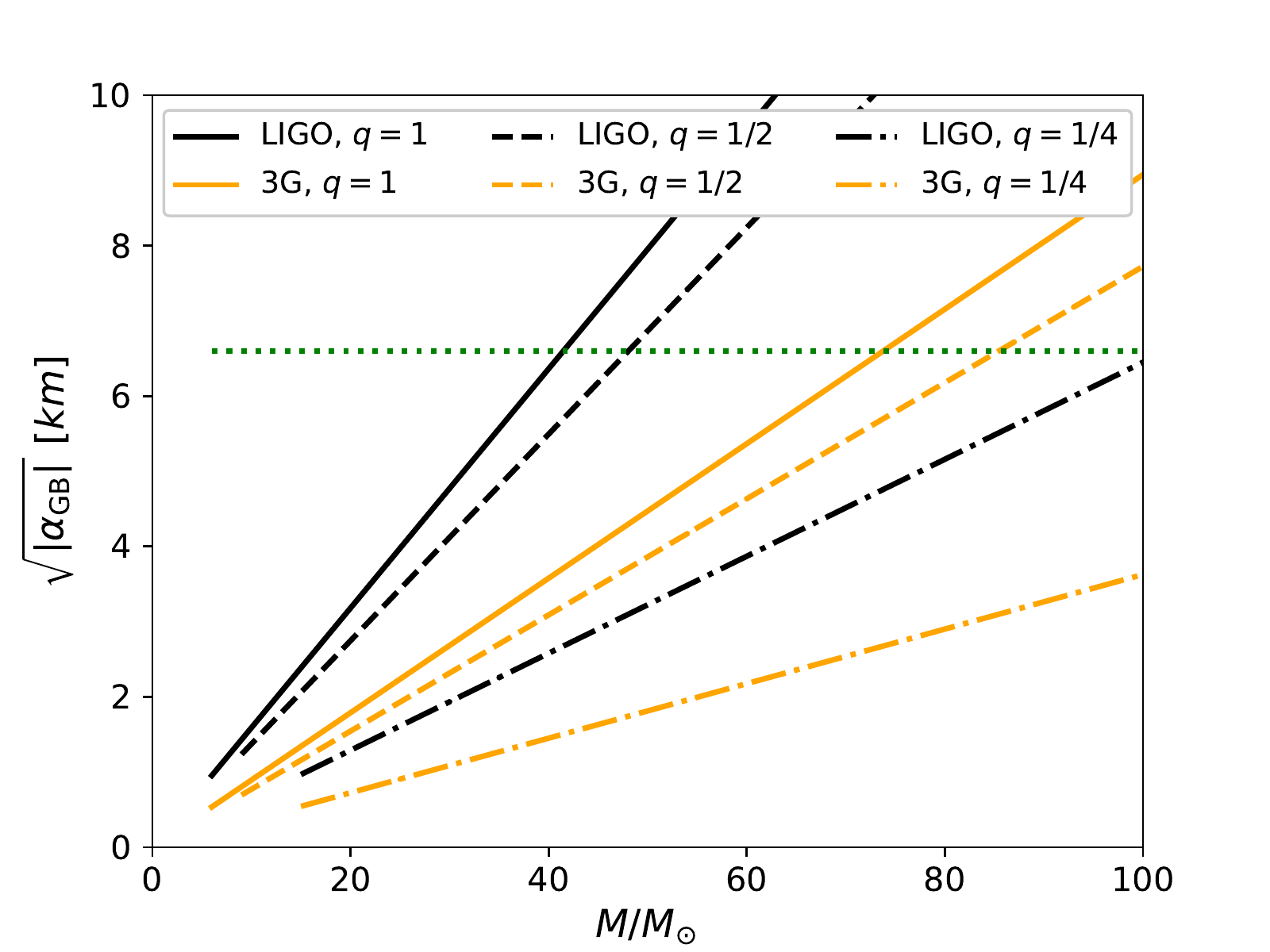}
\caption{\label{fig:BBH_AlpDet_VarM_allq} 
(Color online) Bounds on dimensionful coupling parameter $\alpha_{\rm{GB}} = 4 \epsilon M^{2}$, inferred from the
  detector bound~\eqref{eq:BoundEpsNonDet} and estimated near the merger, for a range of source masses $M=m_{1}
  \frac{q+1}{q}$. We start with the smaller BH's mass at $m_{1}=3M_{\odot}$.
We present estimates for LIGO's O2 run (black lines) and forecasts for third generation ground-based detectors (orange
lines) for all simulated mass ratios $q=1$ (solid lines), $q=1/2$ (dashed lines) and $q=1/4$ (dashed-dotted lines).
For comparison we also show the so far most stringent bound coming from the existence of light BHs (green dotted line).
}
\end{center}
\end{figure}

Finally, in Fig.~\ref{fig:BBH_AlpDet_VarM_allq} we show how the absence of a dephasing from LIGO's O2 run and from third
generation ground-based detectors would constrain the dimensionful coupling parameter $\alpha_{\rm{GB}}$ for a range of
possible source masses.  Let us emphasize that these observational bounds are already stronger than previous ones and
can cover the entire range of total masses up to $M\sim100M_{\odot}$ for mass ratios $q=1/4$.


\begin{table}[htpb!]
\caption{\label{tab:ConstraintsForecast} (Color online) Constraints on the GB coupling constant using the
  estimate~\eqref{eq:BoundEpsNonDet} due to the non-detection of any phase deviation in LIGO events and forecasts for 3G
  and LISA.  We present both the dimensionless and the dimensionful couplings where we assume representative sources,
  namely $M=20M_{\odot}$ and $M=60M_{\odot}$ for ground-based detectors and $M=10^{5}M_{\odot}$ for LISA.  }
\begin{tabular}{|l|c|c|c|c|}
\hline
                                     & $M/M_{\odot}$ & $q=1$  & $q=1/2$ & $q=1/4$       \\
\hline\hline
$10^{4}\, \epsilon_{\rm{LIGO}}$      &               & $28.9$ & $21.6$  & $4.8$         \\
$\sqrt{\alpha_{\rm{GB,LIGO}} }$ [km] & $20$          & $3.2$  & $2.7$   & $1.3$         \\
                                     & $60$          & $9.5$  & $8.2$   & $3.9$         \\
\hline
$10^{4}\,\epsilon_{\rm{3G}}$         &               & $9.2$  & $6.8$   & $1.5$         \\
$\sqrt{\alpha_{\rm{GB,3G}} }$ [km]   & $20$          & $1.8$  & $1.5$   & $0.7$         \\
                                     & $60$          & $5.4$  & $4.6$   & $2.2$         \\
\hline
$10^{4}\,\epsilon_{\rm{LISA}}$       &               & $9.2$  & $6.8$   & $1.5$         \\
$\sqrt{\alpha_{\rm{GB,LISA}} }$ [km] & $10^{5}$      & $8940$ & $7720$  & $3630$        \\
\hline
\end{tabular}

\end{table}

\section{Summary}
In this paper we have considered spinning isolated BHs and binary BH systems in sGB gravity, described by the 
action~\eqref{eq:ActionEdGB}. 
We have worked perturbatively in the coupling constant $\alpha_{\rm{GB}}$ but  nonperturbatively in the fields.
We have solved the field equations to first order in $\alpha_{\rm{GB}}$ and computed the right-hand side of the modified Einstein 
equations to second order. This allowed us to simulate, for the first time, the dynamics of the scalar field in the highly dynamical
strong field regime probed by binary BHs, to obtain complete scalar waveforms 
in this context, and to check the range of validity of our scheme.

We first investigated the formation of scalar hair around a Kerr BH with arbitrary
spin and studied the scalar QNM ringing. The latter contains at least two dominant modes, corresponding to the scalar
and gravitational QNMs of a Kerr BH. We expect the same result to hold qualitatively at second order as well. That is,
we expect that the post-merger ringdown waveform from a BH binary coalescence in sGB gravity will contain both
gravitational-led and scalar-led QNMs.

We then investigated the emission of scalar waves from the coalescence of nonspinning BH binaries with various mass
ratios. While the scalar radiation generally displays the typical chirp signal, the dipole mode (radiated only for
unequal mass binaries) displays a more peculiar behavior and no chirping. This suggests that the scalar field exhibits
interesting dynamics in the pre-merger phase.

In all of our simulations, the axisymmetric components of the scalar field approach the profile of the stationary hairy
BH. This BH remnant is characterized by the mass and  spin only (primary hair) and all 
other multipole moments, 
including the scalar charge (secondary hair), can be written in terms of those. Any deviation from this stationary 
multipolar structure is radiated away during the merger and ringdown, just like in GR, and there is a unique scalar 
configuration that acts an the endpoint of dynamical evolution for a BH of given 
mass and spin. This is a significant generalization of similar results obtained previously~\cite{Benkel:2016kcq, 
Benkel:2016rlz} for spherically symmetric spacetimes.

We calculated the scalar energy flux emitted in our binary simulations. This flux leads to modifications 
of the binary evolutions with respect to GR that are imprinted on the standard tensor modes as dephasing. The effect is of second
order in the coupling so we could not compute it directly.
However, the calculation of the flux allowed us to estimate this effect and derive some qualitative observational bounds
on the sGB coupling constant. Requiring that the dephasing due to the GB corrections is smaller than LIGO's phase
sensitivity, our results imply the preliminary constraint $\sqrt{\alpha_{\rm{GB}}}\lesssim \,2.7$km on the sGB coupling
constant for a GW151226-type source, {\em i.e.}, a BH binary with mass ratio $q=1/2$ and total mass
$M=20M_{\odot}$.  The detection of the GW151226 event does enforce this (approximate) bound, which
may be viewed as the
strongest constraint to date on the coupling constant of sGB gravity. 

Our results indicate that systems with smaller mass ratios and smaller total mass can put even more stringent bounds.
Future third-generation detectors will improve such a constraint by at least a 
factor $\sim3$; taking into account the
fact that 3G will detect several sources, possibly with a larger distribution of mass ratios, the improvement may reach
one order of magnitude.  On the other hand, projective bounds on $\alpha_{\rm GB}$ based on dephasing from 
the future space-based LISA detector are unlikely be competitive: even 
though LISA will provide the most stringent constraints on
the dimensionless coupling constant, $\epsilon=\alpha_{\rm GB}/(4M^2)$, it will probe larger total masses/ curvatures
and this will significantly weaken the constraints on $\alpha_{\rm GB}$ itself. 

It should be stressed that dephasing due to dipolar emission is not the only 
way to constrain $\alpha_{\rm GB}$. 
In fact, the dephasing constraint  relies only on energy loss. One expects to obtain significantly stronger bounds by 
attempting to fit complete inspiral-merger-ringdown waveforms (when available) to specific events or even by stacking events.

As we have argued in detail, within a perturbative approach in the coupling constant (but not in the fields), the
leading-order effects of a scalar field on BHs and their binaries are driven by the coupling of the scalar to the
Gauss-Bonnet invariant in the absence of parity violations. The coupling to the Pontryagin density comes at the same
order if parity invariance is broken. Hence, our results, potentially combined with those of
Ref.~\cite{Okounkova:2017yby} in Chern-Simons gravity provide the complete scalar dynamics, to linear order in the
coupling constant.

Work on second-order effects and on the metric backreaction is ongoing. Future work will also focus on a more detailed
analysis of the parameter space for BH binaries, on the mode excitation for the remnant BHs with various spin values,
and on a more rigorous analysis on the detectability of sGB corrections in the GW signal from BH coalescences.

\begin{acknowledgments}
We thank K.~Yagi for sharing his notes used in Sec.~\ref{sssec:LargeRadiusApprox} and for useful discussions. 
We also thank M.~Okounkova and L.~Stein for useful discussions and comments.
H.W.~acknowledges financial support provided by the European Union's H2020 research and innovation program under Marie
Sklodowska-Curie grant agreement no. BHstabNL-655360, by the Royal Society University Research Fellowship
UF160547 and the Royal Society Research Grant RGF\textbackslash R1\textbackslash 180073.
She also acknowledges partial support by grants FPA-2016-76005-C2-2-P, AGAUR SGR-2017-754.
P.P.~acknowledges financial support provided under the European Union's H2020 ERC, Starting Grant agreement
no.~DarkGRA--757480.
T.P.S.~acknowledges partial support from the STFC Consolidated Grant No.~ST/P000703/1.
This project has received funding from the European Union's Horizon 2020 research and innovation programme under the
Marie Sklodowska-Curie grant agreement No 690904.
The authors would like to acknowledge networking support by the COST Action CA16104.
We thankfully acknowledge the computer resources at Marenostrum IV, Finis Terrae II and LaPalma and the technical
support provided by the Barcelona Supercomputing Center via the PRACE grant Tier-0 PPFPWG, and via the BSC/RES grants
AECT-2017-2-0011, AECT-2017-3-0009 and AECT-2018-1-0014.  The code developed for this paper is based on the
\ETK~\cite{EinsteinToolkit:web} and public at~\cite{ZilhaoWitekCanudaRepository}.  The {\textsc{xTensor}} package for
    {\textsc{mathematica}}~\cite{xAct:web,Brizuela:2008ra} has been used.
\end{acknowledgments}

\appendix
\section{Conventions for coupling constant}~\label{app:CouplingConventions}
Here we summarize various conventions for the normalization of the scalar field and the coupling constant.  While we
follow Ref.~\cite{Kanti:1995vq}, it is useful to compare to the following reference literature.
\begin{itemize}
\item The living review by Yunes \& Siemens~\cite{Yunes:2013dva}, the article of Yagi~et.~al.~\cite{Yagi:2011xp} which
  studied BH binaries in quadratic gravity theories using the PN framework, the review on implications of the first GW
  detections~\cite{Yunes:2016jcc} and the study on observational constraints on EdGB gravity from X-ray
  binaries~\cite{Yagi:2012gp}:
\begin{align}
\label{eq:TrafoCouplingYYSPvsKanti}
  &\kappa_{\rm{YYSP}}      = \frac{1}{\kappa}
  \,,\quad\Phi_{\rm{YYSP}} = \frac{\Phi}{\sqrt{\kappa}}\nonumber\\
  &  \alpha_{\rm{YYSP}}    = \frac{\alpha_{\rm{GB}}}{4\sqrt{\kappa}}\,,\quad
  \mu_{\rm{YYSP}}          = \frac{\mu}{\sqrt{\kappa}}  \,.
\end{align}
These papers consider shift-symmetric gravity, but the bounds are applied to EdGB gravity, since the two theories are
equivalent for weak scalar fields.  Note that in~\cite{Yunes:2013dva,Yagi:2011xp} one should make the further assumption
$\beta=1$.
\item The papers on (no-)hair theorems in shift-symmetric sGB gravity~\cite{Sotiriou:2013qea} and on the formation of
  hairy BHs~\cite{Benkel:2016rlz}:
  \begin{align}
  \frac{M^{2}_{Pl}}{2} = & \frac{1}{\kappa}
  \,,\quad
  \alpha_{\rm{SZ}}     =   \frac{ \alpha_{\rm{GB}}}{4}
  \,.
  \end{align}
\end{itemize}

\section{Known perturbative black hole solutions in Einstein-dilaton Gauss-Bonnet gravity}~\label{app:ana}
In this appendix we present some approximate analytical solutions for isolated and binary BHs in EdGB gravity, which are
used in the main text as a benchmark of our numerical results.

\subsection{Stationary black holes in Einstein-dilaton Gauss-Bonnet gravity}~\label{ssec:BHsolsAna}

\subsubsection{Small-spin approximation}~\label{sssec:SmallSpinApprox}
%
We will benchmark the endstate of the BH coalescence against known rotating solutions within the perturbative approach
in the coupling constant ($\epsilon\ll1$).  In the limit where the spin is small, these solutions are known
analytically~\cite{Pani:2011gy,Ayzenberg:2014aka,Maselli:2015tta}.  Stationary solutions for arbitrary spin and beyond
the perturbative approach have been found numerically with elliptic solvers~\cite{Kleihaus:2015aje}.
Here, we focus on the small spin case and consider solutions up to first order in the
coupling~\cite{Pani:2011gy,Ayzenberg:2014aka} (higher-order analytical solutions are derived in~\cite{Maselli:2015tta})
\begin{align}
\label{eq:SolPhiSmallCouplSmallSpin}
\Phi    = & \epsilon\Phi^{(1)}
        =   \P \frac{M}{r} \left[ 1 + \frac{M}{r} + \frac{4}{3} 
\frac{M^{2}}{r^{2}} \right]
\\ &
          - \P \chi^{2} \frac{M}{4 r} \left[ 1 + \frac{M}{r} + \frac{8}{3} 
\frac{M^{2}}{r^{2}} + 6 \frac{M^{3}}{r^{3}} + \frac{64}{5}\frac{M^{4}}{r^{4}} 
\right]
\nonumber \\ &
          - \P \chi^{2} Y_{20} \frac{28}{15} \sqrt{\frac{\pi}{5}} 
\frac{M^{3}}{r^{3}}
            \left[1 + 3 \frac{M}{r} + \frac{48}{7} \frac{M^{2}}{r^{2}} \right] 
\,,\nonumber
\end{align}
where $\chi$ is the dimensionless spin, $\P = \tfrac{\alpha_{\rm{GB}} }{2M^{2} } = 2 \epsilon$ is the (dimensionless)
scalar charge, and $Y_{20} = \sqrt{\frac{5}{16\pi}} \left[3 \cos^{2}\theta - 1 \right]$ is the $l=2,m=0$ spherical
harmonic.
In the nonspinning case, Eq.~\eqref{eq:SolPhiSmallCouplSmallSpin} reduces to the solutions constructed
in~\cite{Sotiriou:2013qea,Sotiriou:2014pfa} if we replace the dimensionless scalar charge $\P$ with the (dimensionful)
charge $\mu = \alpha_{\rm{GB}} /(2M) = M \P$.

\subsubsection{Arbitrary spin and large-distance approximation}~\label{sssec:LargeRadiusApprox}
Complementary to the previous 
approximate solution, Yagi~\cite{Yunes:2016jcc,Berti:2018cxi,PrivCommKentYagi}
derived an analytic solution (again within the perturbative framework) for rotating BHs with arbitrary spin, which is
valid at large distances, {\em i.e.} $r\gg M$.
In this case the scalar field profile reads
\begin{align}
\label{eq:SolPhiArbSpinLargeR}
\Phi = & \epsilon \Phi^{(1)} 
\nonumber \\
     = & \sum_{l\geq0,{\rm{even}} } \P_{l} \left[\frac{M}{r}\right]^{l+1} 
P_{l}\left(\cos\theta\right)
         \left[ 1 + \mathcal{O} \left(\frac{M}{r}\right) \right]
\,,
\end{align}
where $P_{l}\left(\cos\theta\right)$ denotes the Legendre polynomial, and the lowest lying scalar charge multipoles
$\P_{l}$ are
\begin{subequations}
\label{eq:SolPhiArbSpinLargeRCharges}
\begin{align}
\P_{0} = & \epsilon\, \frac{\chi^{2}-1 + \sqrt{1-\chi^{2} } } {\chi^{2}}
\,,\\
\P_{2} = & - \frac{\epsilon}{3\chi^{2} } \left[
                \sqrt{1-\chi^{2}} \left(2\chi^{2} - 5 \right)
                + 8 - 4\chi^{2} + 2 \chi^{4}      \right]
\nonumber \\ &
        - \frac{2 \epsilon}{\chi^{3} }
          \arctan \left[\frac{\sqrt{1-\chi^{2}}-1 }{\chi} \right]
\,.
\end{align}
\end{subequations}
Taking the $r\to\infty$ limit of~\eqref{eq:SolPhiSmallCouplSmallSpin} and the small spin limit
of~\eqref{eq:SolPhiArbSpinLargeR} both solutions agree. Furthermore, note that the monopole scalar charge $\P_{0}$
reduces to $\P$ up to linear order in the spin, as expected.

\subsection{Post-Newtonian expansion for quasi-circular inspiral}~\label{ssec:PNExpansion}
The leading-order PN scalar waveform in EdGB gravity has been computed in Ref.~\cite{Yagi:2011xp}. We summarize here the
main results and give the explicit expressions for the first radiative multipole moments.

The object trajectories can be parametrized by
\begin{subequations}
  \begin{align}
 \mathbf{x}_1 &= x_1^i=\frac{m_2}{M}b[\cos\omega t,\sin\omega t,0]\,,\\
 \mathbf{x}_2 &= x_2^i=-\frac{m_1}{M}b[\cos\omega t,\sin\omega t,0]\,,
  \end{align}
\end{subequations}
where $M=m_1+m_2$ is the total mass, $b$ and $\omega$ are the orbital distance and frequency. To leading (Newtonian)
order, $\omega=\sqrt{M/b^3}$ and the orbital velocity is $v=\sqrt{M/b}$. We also define $\mathbf{n}= (x_1^i-x_2^i)/b$.

The leading-order solution for the scalar field is
\begin{align}
 \Phi& =\frac{1}{r}\sum_m \frac{1}{m!}\frac{\partial^m}{\partial t^m} 
\int_{\cal M} \mu_1 \delta^{(3)}(\mathbf{x}'-\mathbf{x_1})(\mathbf{n}\cdot 
\mathbf{x'})^m d^3 x'\nonumber\\
& + 1 \leftrightarrow 2\,,
\end{align}
where ${\mu}_i={\alpha_{\rm GB}}/{(2m_i)}$ is the charge parameter of the $i$-th body and we used standard Cartesian
coordinates. Thus, we obtain
\begin{align}
 \Phi &=\sum_m \Phi_m\nonumber\\
 &=\frac{1}{r}\sum_m \frac{1}{m!}\frac{\partial^m}{\partial 
t^m} \left( \mu_1 
(\mathbf{n}\cdot \mathbf{x_1})^m +\mu_2 (\mathbf{n}\cdot 
\mathbf{x_2})^m\right)\,.
\end{align}
In flat-space polar coordinates~\eqref{eq:TrafoSphVsCart}
the various contributions read
\begin{widetext}
\begin{align}
 \Phi_{0} &= \frac{M}{m_1m_2} \frac{\alpha_{\rm GB}}{2R}\,, \\
 \Phi_{1} &= \left(\frac{b(m_2-m_1)}{m_1 
m_2}\omega\sin\theta\sin(\varphi-\omega t)\right) \frac{\alpha_{\rm GB}}{2R}\,, 
\\
 \Phi_{2} &= -\left(\frac{b^2 \omega ^2 \sin ^2(\theta ) 
\left({m_1}^2-{m_1} {m_2}+{m_2}^2\right) \cos (2 \varphi -2 t \omega )}{{m_1} 
{m_2} M}\right) \frac{\alpha_{\rm GB}}{2R}\,, \\
 \Phi_{3} &= \left(\frac{b^3 \omega ^3 \sin ^3(\theta ) ({m_1}-{m_2}) 
\left({m_1}^2+{m_2}^2\right) (\sin (\varphi -t \omega )+9 \sin (3 \varphi -3 t 
\omega ))}{8 {m_1} {m_2} M^2}\right) \frac{\alpha_{\rm GB}}{2R}\,, 
\\
 \Phi_{4} &= \left(\frac{ b^4 \omega ^4 \sin ^4(\theta ) 
\left({m_1}^4-{m_1}^3 {m_2}+{m_1}^2 {m_2}^2-{m_1} {m_2}^3+{m_2}^4\right) (\cos 
(2 \varphi -2 t \omega )+4 \cos (4 \varphi -4 t \omega ))}{3 {m_1} {m_2} 
M^3}\right) \frac{\alpha_{\rm GB}}{2R}\,.
\end{align}
\end{widetext}
The contributions of $m=0$ and $m=1$ agree with those given in Ref.~\cite{Yagi:2011xp}, whereas we explicitly present
also the other multipoles that are relevant in our case. As expected, in the equal-mass case only the even multipoles
are nonvanishing.

From Eq.~\eqref{eq:DefPhilm}, it is easy to check that the \emph{leading-order} contribution to the radiative mode
$\Phi_{ll}$ comes only from $\Phi_{m=l}$, {\em i.e.}
\begin{align}
\label{eq:DefPhilm2}
\Phi_{mm} (t,\rex) = & \int\dif\Omega\, \Phi_{m} 
Y^{\ast}_{mm}(\theta,\varphi)\,.
\end{align}

Finally, the orbital parameters needs to be evolved adiabatically, {\em i.e.} $b\to b(t)$ and $\omega\to\omega(t)$,
where $\omega(t)$ is extracted from the evolution at zeroth order ({\em i.e.}, the GR coalescence).

\bibliographystyle{apsrev4-1}
\bibliography{RefsBBHinEDGB.bib}

\end{document}